\documentclass[onecolumn,aps,prd,preprintnumbers,amsmath,amssymb,showkeys]{revtex4-2}
\usepackage{natbib}
\usepackage{graphicx}
\usepackage{bm}
\usepackage{color}
\usepackage{xcolor}
\usepackage{amsmath}
\usepackage{amsfonts}
\usepackage{amssymb}
\usepackage{makeidx}
\usepackage{subcaption}
\usepackage{physics}
\usepackage{hyperref}
\usepackage{array}
\usepackage{multirow}%
\usepackage{quantikz}
\usetikzlibrary{quantikz2}
\usepackage{float}
\usepackage{algorithm}%
\usepackage{algorithmicx}%
\usepackage{tikz}
\usepackage{graphicx}
\usepackage{subcaption}
\usepackage{mathtools}
\begin{document}
	\title[Article Title]{Environment-assisted and weak measurement strategies for robust bidirectional quantum teleportation}
	\author{Javid Ahmad Malik$^1$, 
		Muzaffar Qadir Lone$^2$\footnote{Corresponding author: lone.muzaffar@uok.edu.in}, 
		Prince A Ganai$^1$}
	\affiliation{$^1$Department of Physics, National Institute of Technology, Srinagar-190006 India\\
		$^2$Quantum Dynamics Lab, Department of Physics, University of Kashmir, Srinagar-190006 India}

	\begin{abstract}
		
		This paper presents strategies for enhancing the robustness of bidirectional quantum teleportation (BQT) through environment-assisted and weak measurement techniques. BQT is a crucial component of distributed quantum networks, allowing for the bilateral transfer of quantum information between two nodes. While perfect teleportation necessitates maximally entangled states, these are vulnerable to degradation due to inherent decoherence. We propose a BQT scheme that enables the bilateral transfer of arbitrary qubits between nodes via amplitude damping channels (ADC), aiming to optimize fidelity using weak measurements in the final step of the process. Environment-assisted measurements (EAM) are used to establish a four-qubit channel composed of two Bell states. We explore two situations: (I) where only the recovery qubits pass through amplitude damping channels and (II) where the entire four-qubit channel is subjected to ADC. Our findings demonstrate a balance between average fidelity and success probability when the weak measurement strength ($q_w$) is constrained by the decay rate ($p$), specifically $q_w \in {[0,p]}$. Perfect BQT is achieved when $q_w = p$, indicating complete suppression of ADC effects. On the other hand, a decline in both average fidelity and success probability is noted when the weak measurement strength surpasses the ADC strength, marking the prohibited domain as $q_w \in {(p,1]}$. Additionally, our secured BQT protocol consistently outperforms the unprotected scheme in both scenarios, highlighting the effectiveness of the proposed protection strategies.
		
	\end{abstract}
\maketitle
	\section{Introduction}\label{sec1}
Quantum entanglement is the most striking and counter intuitive feature of quantum mechanics that  enables us to perform computational tasks which are classically impossible\cite{horo}. Entanglement acts as a resource for many quantum information tasks such as quantum teleportation \cite{ch}, superdense coding\cite{qsd}, quantum cryptography\cite{qc}, quantum state sharing\cite{qss}, etc. 
Quantum teleportation is a technique to transfer quantum information between parties exploiting a distributed entangled state and a classical communications channel\cite{ch}.
After the introduction of the original teleportation protocol by Bennett\textit{et al.}\cite{ch}, multi-qubit entangled states have been utilized to achieve several variants of quantum teleportation protocol like bidirectional quantum teleportation (BQT)\cite{bqt1,bqt2,bqt3,bqt4}, bidirectional controlled quantum teleportation (BCQT) \cite{bqct1,bqct2,bqct3,bqct4,bqct5} etc. By enabling the simultaneous exchange of quantum states between two nodes, BQT optimizes the use of network resources and reduces latency, which is crucial for maintaining the coherence and integrity of quantum information over long distances\cite{bqtad}.   Furthermore, in a quantum network, BQT can support distributed quantum computing by allowing quantum information to be shared and processed across different nodes. In BCQT, the teleportation process is supervised by the third party who prepares the entangled channel and distributes among the two partners adding an additional layer of security to the protocol\cite{bqct5}.

Many of the studies mentioned above rely on oversimplified and idealized assumptions, including the absence of noise and decoherence. These studies frequently employ maximally entangled channels without accounting for interactions between the environment and the system, which are unavoidable and significantly diminish the degree of entanglement \cite{deco0,deco1,deco2,deco3,deco4,deco5}. Consequently, this can result in information loss \cite{nqt1,nqt2,nqt3,nqt4,nbqt1}. To counteract the effects of noise, different strategies have been implemented with varying degrees of complexity and effectiveness. One of the methods to overcome the noise involves modifying the unitary operations applied by the receiver in  quantum teleportation protocol to counteract specific noise effects\cite{pro1,pro2,pro3,pro4}. Another strategy is entanglement purification which aims to boost the fidelity of teleportation in noisy environments by enhancing the entanglement of a  pair of qubits\cite{enpro1,enpro2,enpro3,enpro4}.
 Quantum error correction is one of the key techniques used to protect qubits from noise when they are transmitted through noisy channels. This method involves encoding quantum information into a larger Hilbert space using additional qubits. This need for extra qubits (ancilla), increases the complexity and resource requirements of quantum systems\cite{qec1,qec2,qec3,qec4,qec5}. To reduce complexity and resource usage, strategies utilizing weak measurements to counteract decoherence have been proposed theoretically\cite{twm1,twm4,twm5,twm6} and implemented experimentally\cite{wme1,wme2,wme3,wme4,wme5,wme6}.
  Weak measurement can mitigate the perturbation of a quantum system by attenuating the interaction responsible for the measurement process. A weak measurement yields reduced information about the system while minimizing its disturbance and forms the basis of our optimized BQT protocol in this paper. 
 
  The above protection schemes manipulate the state of the system and optimizes the teleportation. In addition, another powerful technique known as environment-assisted measurement (EAM) is used to enhance teleportation fidelity \cite{eam1,eam2,eam3,eam4}. This technique involves the manipulation of noisy channel during entanglement distribution by applying measurement on the environment\cite{eamr1,eamr2}. In the presence of noise, the evolution of the system is described by the expression $ \sum_n K_n \rho_S(0) K_n^\dagger $, where $ K_n $ are the Kraus operators representing the environment, and $ \rho_S(0) $ is the initial state of the system. EAM leads the environment to collapse into its respective eigenstate for each $ n $-th observation outcome. Consequently, the system is projected into the state $ \rho_S^{(n)} = K_n \rho_S(0) K_n^\dagger $. This method involves selecting outcomes associated with reversible Kraus operators and disregarding all others. At the end by applying weak measurement reversal to the system, the initial state can be restored. All these protection schemes aim to improve unidirectional quantum teleportation. However, no similar protection scheme has been proposed for BQT. Since BQT serves as the foundation for distributed quantum networks by enabling pairwise information flow between quantum nodes, it is crucial to adapt similar techniques to optimize BQT against decoherence, a goal that is addressed in this paper.

Motivated by the use of environment-assisted measurement (EAM) equipped with weak measurements we propose protection strategies for optimized BQT. In this paper, we have discussed the protection against amplitude damping noise alone. However, the protocol works well for decoherence channels with a minimum of one invertible Kraus operator. To increase the security  a third party Charlie supervises the protocol during entanglement distribution.  We analyze two scenarios based on the mode of entanglement distribution by a third party. In first case only recovery qubits are exposed to noise. The second case involves passing all the qubits through independent noisy channels having same decay rate. In both the cases, recipients apply EAM on the noisy channel and consider the result corresponding to the ground state, while the result for excited state is discarded. Once the recipients communicate the EAM results to supervisor  the quantum channel between the two partners is successfully established and  the BQT protocol begins. The final step of the BQT involves application of the unitary operators equipped with specially designed weak measurements operators to suppress the decoherence.
We adjust the weak measurement strength, which can be externally controlled, to evaluate its effectiveness in counteracting decoherence by examining average fidelity and teleportation success probability. It is observed by setting a balance between noise and protection the BQT is achieved with unit fidelity even in presence of noise. Also, we have identified distinct domains of weak measurement strength: one that optimizes the protocol and another that undermines it, making it crucial to carefully tune the measurement strength. By switching  off  the weak measurements it is observed that EAM are effective in channel protection. Furthermore, a remarkable enhancement in average fidelity is observed over unprotected BQT protocol marking the novelty of our optimization strategies.
The procedure for the proposed optimized BQT protocol involves several key steps: first, a third party prepares two Bell states and distributes them through noisy channels. Both recipients then apply EAM to the qubits affected by noise, resulting in the creation of an entangled channel between them. The BQT process is initiated thereafter, and finally, weak measurements complete the  protocol.

The remainder of the paper is organized as follows: Section \ref{sub1} introduces the proposed protected BQT protocol and discusses the first scenario where only the recovery qubits are exposed to noise. In Section \ref{sub2}, we explore the second scenario, in which all channel qubits are subjected to noise. These sections also provide a detailed numerical and graphical analysis of the average fidelity and total success probability, the primary performance indicators used in this study. Finally, our conclusions are presented in Section \ref{sec3}.

\section{BQT with Amplitude Damping on Recovery Qubits ($2$ and $3$)}
	\label{sub1}
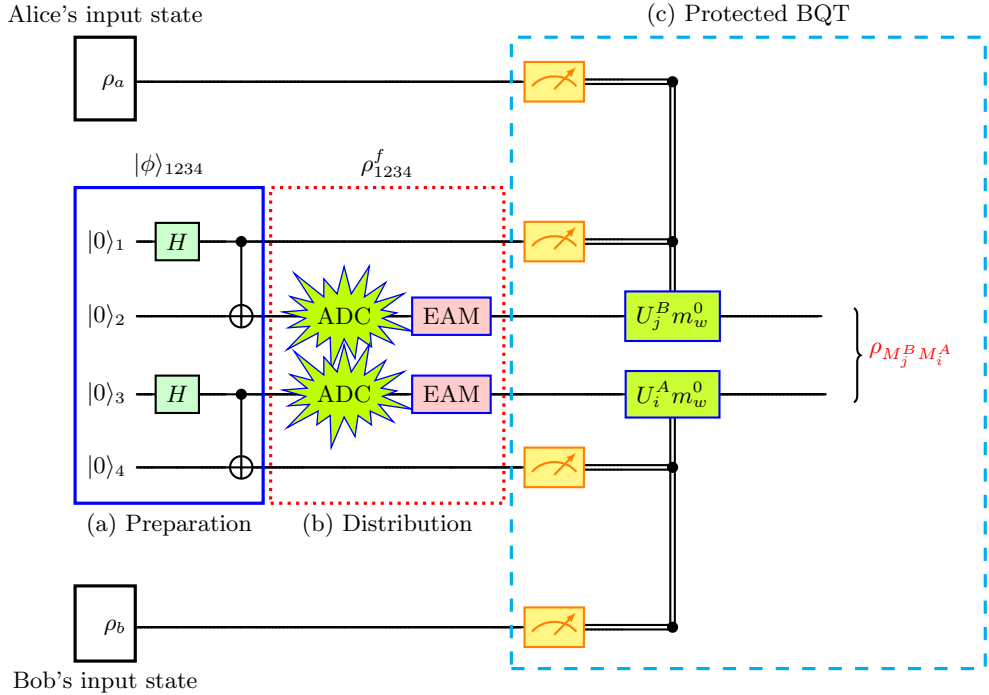
\begin{figure}[H]
		\centering
		\tikzset{
			noisy/.style={
				starburst,
				fill=lime,
				draw=blue,
				line width=.5pt, % Make the starburst border thicker
				inner xsep=-4pt,
				inner ysep=-1pt
			},
			quantum wire/.style={thick}, % Make wires thicker
			gate/.style={draw, thick, minimum size=3em}, % Make gates bold
			label style/.style={font=\bfseries\boldmath} % Bold labels
		}
		\begin{quantikz}[row sep=4mm, column sep=0.5mm]
			\lstick{$\rho_{a}$}&\qw&\qw&\qw&\qw&\qw&\qw&\qw&\qw&\qw&\qw&\qw&\qw&\qw&\qw&\qw&\qw&\qw&\qw&\qw&\qw&\qw&\qw&\qw&\qw&\qw&\qw&\qw&\qw&\qw&\qw&\qw&\qw&\qw&\qw&\qw&\qw&\qw&\qw&\qw&\qw&\qw&\qw&\qw&\meter[style={fill=yellow!60, draw=orange, thick}]{}-
			&\setwiretype{c}&&&&&&&&&&\ctrl[vertical
			wire=c]{5}\\\\\\\\
			\lstick{$|0\rangle_1$}&\qw&\qw &\qw&\qw&\gate[style={draw=black, fill=green!20}]{H}&\qw&\qw&\qw&\qw&\qw&\qw&\qw&\ctrl{1}&\qw&\qw&\qw&\qw&\qw&\qw&\qw&\qw&\qw&\qw&\qw&\qw&\qw&\qw&\qw&\qw&\qw&\qw&\qw&\qw&\qw&\qw&\qw&\qw&\qw&\qw&\qw&\qw&\qw&\qw&\meter[style={fill=yellow!60, draw=orange, thick}]{}-
			&\setwiretype{c}&&&&&&&&&&\control\cw\\
			\lstick{$|0\rangle_2$}&\qw&\qw&\qw&\qw&\qw&\qw&\qw&\qw&\qw&\qw&\qw&\qw&\targ{}&\qw&\qw&\qw&\qw&\qw&\qw&\qw&\qw&\qw&\qw&\qw&\qw&\qw&\gate[1,style={noisy},label
			style=black]{\text{ADC}}&\qw&\qw&\qw&\qw&\qw&\qw&\qw&\gate[1,style={fill=pink!80, draw=blue, label style={text=red}}]{\text{EAM}}&\qw&\qw&\qw&\qw&\qw&\qw&\qw&\qw&\qw&\qw&\qw&\qw&\qw&\qw&\qw&\qw&\qw&\qw&\qw&\gate[1,style={fill=lime!80, draw=blue, label style={text=red}}]{\text{$U_j^B m_{w}^0$}}&\qw&\qw&\qw&\qw&\qw
			&\qw&\qw&\qw&\qw&\qw&\qw&\qw&\qw&\qw&\qw&\qw&\qw&\qw&\qw&\qw&\qw&\qw&\qw&\qw&\qw&\qw&~~~ \rstick[2]{\textcolor{red}{$~~~\rho_{M_j^B M_i^A}$}}  
			\\
			\lstick{$|0\rangle_3$}&\qw&\qw &\qw&\qw&\gate[style={draw=black, fill=green!20}]{H}&\qw&\qw&\qw&\qw&\qw&\qw&\qw&\ctrl{1}&\qw&\qw&\qw&\qw&\qw&\qw&\qw&\qw&\qw&\qw&\qw&\qw&\qw&\gate[1,style={noisy},label
			style=black]{\text{ADC}}&\qw&\qw&\qw&\qw&\qw&\qw&\qw&\gate[1,style={fill=pink!80, draw=blue, label style={text=red}}]{\text{EAM}}&\qw&\qw&\qw&\qw&\qw&\qw&\qw&\qw&\qw&\qw&\qw&\qw&\qw&\qw&\qw&\qw&\qw&\qw&\qw&\gate[1,style={fill=lime!80, draw=blue, label style={text=red}}]{\text{$U_i^A m_{w}^0$}}
			&\qw&\qw&\qw&\qw&\qw
			&\qw&\qw&\qw&\qw&\qw&\qw&\qw&\qw&\qw&\qw&\qw&\qw&\qw&\qw&\qw&\qw&\qw&\qw&\qw&\qw&\qw&\qw~~\\
			\lstick{$|0\rangle_4$}&\qw&\qw&\qw&\qw&\qw&\qw&\qw&\qw&\qw&\qw&\qw&\qw&\targ{}&\qw&\qw&\qw&\qw&\qw&\qw&\qw&\qw&\qw&\qw&\qw&\qw&\qw&\qw&\qw&\qw&\qw&\qw&\qw&\qw&\qw&\qw&\qw&\qw&\qw&\qw&\qw&\qw&\qw&\qw&\meter[style={fill=yellow!60, draw=orange, thick}]{}-
			&\setwiretype{c}&&&&&&&&&&\control\cw\\\\\\\\
			\lstick{$\rho_{b}$}&\qw&\qw&\qw&\qw&\qw&\qw&\qw&\qw&\qw&\qw&\qw&\qw&\qw&\qw&\qw&\qw&\qw&\qw&\qw&\qw&\qw&\qw&\qw&\qw&\qw&\qw&\qw&\qw&\qw&\qw&\qw&\qw&\qw&\qw&\qw&\qw&\qw&\qw&\qw&\qw&\qw&\qw&\qw&\meter[style={fill=yellow!60, draw=orange, thick}]{}-
			&\setwiretype{c}&&&&&&&&&&\ctrl[vertical
			wire=c]{-5}\\	
		\end{quantikz}
		\begin{tikzpicture}[overlay]
			\node[draw=black,very thick, fit={(-11.8,3.5) (-11.2,4.4)}, inner sep=1mm, label=above:{ Alice's input state}] {};
			\node[draw=blue,very thick, fit={(-11.8,2.4) (-9.5,-1.6)}, inner sep=1mm, label=below:{(a) Preparation}, label=above:{ $| \phi\rangle_{1234}$}] {};
			\node[draw=red,very thick,dotted, fit={(-9.2,-1.6) (-6.3,2.4)}, inner sep=1mm,label=below:{(b) Distribution} ,label=above:{ $ \rho^f_{1234}$}] {};
			\node[draw=cyan, very thick, dashed, dash pattern=on 5pt off 5pt, fit={(0.1,-3.8) (-6,4.4)}, inner sep=1mm,label=above:{(c) Protected BQT}] {};
			\node[draw=black,very thick, fit={(-11.8,-2.9) (-11.2,-3.7)}, inner sep=1mm, label=below:{ Bob's input state}] {};
		\end{tikzpicture}
		\caption{ Quantum circuit of protected BQT  when only recovery qubits ($2$ and $3$) are passed through ADCs. (a) Four qubit channel preparation Eq.\ref{eqtch1} (b) Entanglement distribution by third party  (c) Protected BQT wherein weak measurements are applied. $U_i^A m_w^0$ and $U_j^B m_w^0$ are the unitary operators of Alice and Bob respectively, equipped with weak measurements  and $\rho_{M^{B}_j M^{A}_i}$ the teleported state post protection.}
		\label{rbqt}
	\end{figure}

	In this section, we introduce the BQT scheme in presence of noise (Amplitude damping) using two bell-states. Weak measurement and EAM strategies are applied to optimize the BQT protocol for enhanced fidelity. A third party, Charlie, prepares two Bell states $|\phi^+\rangle_{12} \otimes |\phi^+\rangle_{34}$, and distributes the qubits such that qubits $1$ and $3$ are sent to Alice, while qubits $2$ and $4$ are sent to Bob as shown in Figs.\ref{rbqt} and \ref{ebqt}. The two partners apply EAM on the qubits subjected to noise and communicate the measurement outcomes to Charlie. This process establishes an entangled channel between Alice and Bob. After this the BQT protocol begins and in the final step both partners apply weak measurements on qubits $2$ and $3$ to mitigate the decoherence. Here, the first scenario where only recovery qubits $2$ and $3$ are passed through independent ADCs with same decay rate $p$ is discussed in detail. The quantum circuit for the same  is shown in Fig.\ref{rbqt}.
	 
	 The four qubit pure entangled state $|\phi\rangle_{1234}$, prepared by charlie from single qubit states $|0\rangle$ as shown in Fig.\ref{rbqt} is given by
\begin{eqnarray}
	|\phi\rangle_{1234}= \frac{1}{2}\big(|0000\rangle+|0011\rangle+|1100\rangle+|1111\rangle\big)_{1234}
	\label{eqtch1}
\end{eqnarray} 
and the corresponding density operator is 
\begin{eqnarray}
	 \rho_{1234}=|\phi\rangle_{1234}\langle\phi|
	 \label{eqd}
\end{eqnarray} 
Now, the third party distributes qubits $(1,3)$ to Alice and $(2,4)$ to Bob, where only qubits $3$ and $2$ are transmitted through ADCs. The noisy channels are characterized by the following Kraus operators:

\begin{equation}
k_0 =
\begin{bmatrix}
	1&0\\
	0&\sqrt{1-p}\\
	\end{bmatrix},  k_1=
	\begin{bmatrix}
	0&\sqrt{p}\\
	0&0\\
	\end{bmatrix} 
	\label{ko}
\end{equation}
where, $p=1-e^{-\gamma_0t}$ quantifies the damping strength and can be seen as the decay probability from the excited state $|1\rangle$ to ground state $|0\rangle$ with $ \gamma_0$ as the decay rate. Since only qubits ($3$ and $2$) are subjected to ADC, the corresponding four qubit Kraus operators denoted by ($K_i^A$ and $K_j^B$ ) are 
\begin{eqnarray}
	K_i^A = I \otimes I \otimes k_i \otimes I,~~~~~ 
	K_j^B = I \otimes k_j \otimes I \otimes I, \quad (i,j = 0,1)
\end{eqnarray}
Alice and Bob implement EAM on the decoherence channels where 
a measurement is performed on the channel by each party, causing it to collapse into the eigenstates of the measured observable. As a result, the system is projected into a state corresponding to the channel's collapsed state. If the channel collapses into the $n^{th}$ eigenstate ($n = 0, 1$), the system transitions to the state represented by $\rho_S^n = \frac{K_n \rho_S(0) K_n^\dagger}{Tr(K_n \rho_S(0) K_n^\dagger) }$.

For our analysis, we focus on the measurement outcome associated with the invertible Kraus operator \(k_0\) and disregard the outcome corresponding to \(k_1\). Thus when two parties detect the channels to be in an unexcited state $k_0$ and upon conveying the results to Charlie, the entanglement distribution process is completed and subsequently BQT is achieved. Now the corresponding distributed entangled channel is given by
\begin{eqnarray}
	\rho_{1234}^f &=& \frac{K_0^A K_0^B \rho_{1234} K_0^{B\dagger} K_0^{A\dagger}}{\text{Tr}(K_0^A K_0^B \rho_{1234} K_0^{B\dagger} K_0^{A\dagger})} 
\end{eqnarray}
	\begin{equation}
	%\footnotesize
	\rho_{\text{1234}}^f =\frac{1}{4g_{32}^{EAM}} \begin{pmatrix}
	1 & \cdots & \sqrt{1-p} & \cdots &\sqrt{1-p} & \cdots & (1-p) \\
		\vdots & \cdots & \vdots & \cdots & \vdots & \cdots & \vdots \\
		\sqrt{1-p} & \cdots & (1-p) & \cdots & (1-p) & \cdots & (1-p)\sqrt{1-p} \\
		\vdots & \cdots & \vdots & \cdots & \vdots & \cdots & \vdots \\
		\vdots & \cdots & \vdots & \cdots & \vdots & \cdots & \vdots \\
		\sqrt{1-p} & \cdots & (1-p) & \cdots & (1-p)& \cdots & (1-p)\sqrt{1-p} \\
		\vdots & \cdots & \vdots & \cdots & \vdots & \cdots & \vdots \\
		(1-p) & \cdots & (1-p)\sqrt{1-p} & \cdots & (1-p)\sqrt{1-p} & \cdots & (1-p)^2
	\end{pmatrix}
	\label{chem}
\end{equation}
where $g_{32}^{EAM}=\text{Tr}(K_0^A K_0^B \rho_{1234} K_0^{B\dagger} K_0^{A\dagger})=\frac{(2-p)^2}{4}$ is the success probability of entanglement distribution, clearly for $p=0$ that is when noisy effects are ignored the entanglement distribution occurs with unit success probability. Now the BQT protocol is implemented in the following manner (see Fig.~\ref{rbqt}):

 Alice  and Bob are in possession of arbitrary single qubit states given below in Eq.~\ref{inp}, which they intend to teleport bidirectionally 
 
	\begin{equation}
	\rho_A = \begin{pmatrix}
		|\alpha|^2& \alpha\beta^* \\
		\alpha^*\beta &|\beta|^2
	\end{pmatrix},~~~~~  \rho_B = \begin{pmatrix}
	|\gamma|^2& \gamma\delta^* \\
	\gamma^*\delta &|\delta|^2
	\end{pmatrix}
	\label{inp}
\end{equation}
with $\text{Tr}(\rho_A)=\text{Tr}(\rho_B)=1$. They interact their input qubits with their share of the distributed entangled channel Eq.~\ref{chem} and the resulting total state of the six-qubits is $\rho_{t}=\rho_A\otimes \rho_{1234}^f \otimes \rho_B$. Now, Alice and Bob project the qubit pairs $(a,1)$ and $(4,b)$ onto the Bell bases $|\eta_i^A\rangle$ and $|\zeta_j^B\rangle$, respectively, defined as follows:
 
\begin{widetext}
	\begin{minipage}{0.48\textwidth}
		\begin{equation}
			\begin{aligned}
				|\eta_1^A\rangle &= \frac{1}{\sqrt{2}}\big(|00\rangle+|11\rangle\big) \\ 
			|\eta_2^A\rangle  &= \frac{1}{\sqrt{2}}\big(|00\rangle-|11\rangle\big) \\ 
				|\eta_3^A\rangle  &= \frac{1}{\sqrt{2}}\big(|01\rangle+|10\rangle\big) \\ 
				|\eta_4^A\rangle  &= \frac{1}{\sqrt{2}}\big(|01\rangle-|10\rangle\big) \nonumber
			\end{aligned}
		\end{equation}
	\end{minipage}%
	\hspace{-0.1cm}
	\begin{minipage}{0.48\textwidth}
		\begin{equation}
			\begin{aligned}
				|\zeta_1^B\rangle &= \frac{1}{\sqrt{2}}\big(|00\rangle+|11\rangle\big) \\ 
				|\zeta_2^B\rangle  &= \frac{1}{\sqrt{2}}\big(|00\rangle-|11\rangle\big) \\ 
				|\zeta_3^B\rangle  &= \frac{1}{\sqrt{2}}\big(|01\rangle+|10\rangle\big) \\ 
				|\zeta_4^B\rangle &= \frac{1}{\sqrt{2}}\big(|01\rangle-|10\rangle\big) 
			\end{aligned}
			\label{bst}
		\end{equation}
	\end{minipage}
\end{widetext}
the corresponding projection operators  acting on the six qubit total state $\rho_{t}=\rho_a\otimes \rho_{1234}^f \otimes \rho_b$ are 
\begin{eqnarray}
\Pi^A_i = | \eta_i^A \rangle \langle \eta_i^A \otimes I\otimes I\otimes I\otimes I\quad (i=1,2,3,4)
\label{prA}
\end{eqnarray}
\begin{eqnarray}
	\Pi^B_j = I\otimes I\otimes I\otimes I \otimes| \zeta_j^B \rangle \langle \zeta_j^B |
	\quad (j=1,2,3,4)
	\label{prB}
\end{eqnarray}
where Eq.~$\ref{prA}$ acts only on Alice's qubits $(a, 1)$, and Eq.~$\ref{prB}$ on Bob's qubits $(4, b)$, leaving the remaining qubits unaffected. The measurement results are conveyed to each other  classically and the corresponding post-measurement state of the two partners is
\begin{eqnarray}
	\tilde{\rho_t}=\frac{\Pi^A_i\Pi^B_j\rho_{t}\Pi^{B\dagger}_j\Pi^{A\dagger}_i	}{\text{Tr}(\Pi^A_i\Pi^B_j\rho_{t}\Pi^{B\dagger}_j\Pi^{A\dagger}_i)}
\end{eqnarray} 
where $\text{Tr}(\Pi^A_i\Pi^B_j\rho_{t}\Pi^{B\dagger}_j\Pi^{A\dagger}_i)$ is the joint probability of Alice and Bob measuring the bell states $|\eta_i^A\rangle$ and  $|\zeta_j^B\rangle$ defined in Eq.~\ref{bst}. Also, the individual probabilities of the two partners are $\text{P}_i^A=\text{Tr}(\Pi_i^A\rho_t\Pi_i^{A\dagger})$ and $\text{P}_j^B=\text{Tr}(\Pi_j^B\rho_t\Pi_j^{B\dagger})$ respectively. Since during EAM one of the measurement outcome was discarded the BQT becomes probabilistic. The joint probability for each of the sixteen possible combinations is then given by $\text{P}_i^A \text{P}_j^B$.
\begin{eqnarray}
	\text{P}_i^A \text{P}_j^B &=\frac{1}{(4-2p)^2}& 
	\left\{
	\begin{aligned}
		& \big( (1-|\beta|^2p)(1-|\delta|^2p) \big) \quad &\text{for } i,j=1,2 \\
		& \big( (1-|\beta|^2p)(1-|\gamma|^2p) \big) \quad &\text{for } i=1,2 \text{ \& } j=3,4 \\
		& \big( (1-|\alpha|^2p)(1-|\delta|^2p) \big) \quad &\text{for } i=3,4 \text{ \& } j=1,2 \\
		& \big( (1-|\alpha|^2p)(1-|\gamma|^2p) \big) \quad &\text{for } i,j=3,4
	\end{aligned}
	\right.
	\label{pro}
\end{eqnarray}
The recovered states, prior to ADC suppression via weak measurements, are obtained by tracing out the projected qubits $(4,b)$ and $(a,1)$ by Alice and Bob, respectively, as follows:
\begin{eqnarray}
\rho_{23}=\text{Tr}_{a1}\text{Tr}_{4b}\bigg(\frac{\Pi^A_i\Pi^B_j\rho_{t}\Pi^{B\dagger}_j\Pi^{A\dagger}_i	}{\text{Tr}(\Pi^A_i\Pi^B_j\rho_{t}\Pi^{B\dagger}_j\Pi^{A\dagger}_i)}\bigg)
\label{nn}
\end{eqnarray}
 and the projective measurement outcomes of the two partners along with corresponding recovered states are listed in Tab.~\ref{tb1}
\begin{table}[H]
	\centering
	\begin{tabular}{|c|c|c|}
		\hline
		Bob's and Alice's Recovered state ($\rho_{23}$) & \multicolumn{2}{c|}{Measurement Outcomes} \\ 
		\cline{2-3} % Adds a horizontal line below the "Measurement Outcomes" label
		&~~~~ Bob~~~ & Alice \\ 
		\hline
		$\rho_{23} = \frac{1}{(4-2p)^2} 
		\begin{pmatrix}
			|\alpha|^2 & \alpha\beta^*\sqrt{1-p} \\
			\alpha^*\beta\sqrt{1-p} & |\beta|^2(1-p)
		\end{pmatrix} \otimes \begin{pmatrix}
			|\gamma|^2 & \gamma\delta^*\sqrt{1-p} \\
			\gamma^*\delta\sqrt{1-p} & |\delta|^2(1-p)
		\end{pmatrix}$ & $|\zeta_1^B\rangle$ & $|\eta_1^A\rangle$ \\ 
		\hline
			$\rho_{23} = \frac{1}{(4-2p)^2} 
		\begin{pmatrix}
			|\alpha|^2 & -\alpha\beta^*\sqrt{1-p} \\
			-\alpha^*\beta\sqrt{1-p} & |\beta|^2(1-p)
		\end{pmatrix} \otimes \begin{pmatrix}
			|\gamma|^2 & -\gamma\delta^*\sqrt{1-p} \\
			-\gamma^*\delta\sqrt{1-p} & |\delta|^2(1-p)
		\end{pmatrix}$ & $|\zeta_2^B\rangle$ & $|\eta_2^A\rangle$ \\ 
		\hline
			$\rho_{23} = \frac{1}{(4-2p)^2} 
		\begin{pmatrix}
			|\beta|^2 & \beta\alpha^*\sqrt{1-p} \\
			\beta^*\alpha\sqrt{1-p} & |\alpha|^2(1-p)
		\end{pmatrix} \otimes \begin{pmatrix}
			|\delta|^2 & \delta\gamma^*\sqrt{1-p} \\
			\delta^*\gamma\sqrt{1-p} & |\gamma|^2(1-p)
		\end{pmatrix}$ & $|\zeta_3^B\rangle$ & $|\eta_3^A\rangle$ \\ 
		\hline
			$\rho_{23} = \frac{1}{(4-2p)^2} 
		\begin{pmatrix}
			|\beta|^2 & -\beta\alpha^*\sqrt{1-p} \\
			-\beta^*\alpha\sqrt{1-p} & |\alpha|^2(1-p)
		\end{pmatrix} \otimes \begin{pmatrix}
			|\delta|^2 & -\delta\gamma^*\sqrt{1-p} \\
			-\delta^*\gamma\sqrt{1-p} & |\gamma|^2(1-p)
		\end{pmatrix}$ & $|\zeta_4^B\rangle$ & $|\eta_4^A\rangle$ \\ 
		\hline
			$\rho_{23} = \frac{1}{(4-2p)^2} 
		\begin{pmatrix}
			|\alpha|^2 & \alpha\beta^*\sqrt{1-p} \\
			\alpha^*\beta\sqrt{1-p} & |\beta|^2(1-p)
		\end{pmatrix} \otimes \begin{pmatrix}
			|\gamma|^2 & -\gamma\delta^*\sqrt{1-p} \\
			-\gamma^*\delta\sqrt{1-p} & |\delta|^2(1-p)
		\end{pmatrix}$ & $|\zeta_1^B\rangle$ & $|\eta_2^A\rangle$ \\ 
		\hline
			$\rho_{23} = \frac{1}{(4-2p)^2} 
		\begin{pmatrix}
			|\alpha|^2 & -\alpha\beta^*\sqrt{1-p} \\
			-\alpha^*\beta\sqrt{1-p} & |\beta|^2(1-p)
		\end{pmatrix} \otimes \begin{pmatrix}
			|\gamma|^2 & \gamma\delta^*\sqrt{1-p} \\
			\gamma^*\delta\sqrt{1-p} & |\delta|^2(1-p)
		\end{pmatrix}$ & $|\zeta_2^B\rangle$ & $|\eta_1^A\rangle$ \\ 
		\hline
			$\rho_{23} = \frac{1}{(4-2p)^2} 
		\begin{pmatrix}
			|\alpha|^2 & \alpha\beta^*\sqrt{1-p} \\
			\alpha^*\beta\sqrt{1-p} & |\beta|^2(1-p)
		\end{pmatrix} \otimes \begin{pmatrix}
		|\delta|^2 & \delta\gamma^*\sqrt{1-p} \\
		\delta^*\gamma\sqrt{1-p} & |\gamma|^2(1-p)
		\end{pmatrix}$ & $|\zeta_1^B\rangle$ & $|\eta_3^A\rangle$ \\  
		\hline
		$\rho_{23} = \frac{1}{(4-2p)^2} 
		\begin{pmatrix}
			|\alpha|^2 & \alpha\beta^*\sqrt{1-p} \\
			\alpha^*\beta\sqrt{1-p} & |\beta|^2(1-p)
		\end{pmatrix} \otimes \begin{pmatrix}
			|\delta|^2 & -\delta\gamma^*\sqrt{1-p} \\
			-\delta^*\gamma\sqrt{1-p} & |\gamma|^2(1-p)
		\end{pmatrix}$ & $|\zeta_1^B\rangle$ & $|\eta_4^A\rangle$ \\  
		\hline
		$\rho_{23} = \frac{1}{(4-2p)^2} 
		\begin{pmatrix}
			|\alpha|^2 & -\alpha\beta^*\sqrt{1-p} \\
			-\alpha^*\beta\sqrt{1-p} & |\beta|^2(1-p)
		\end{pmatrix} \otimes \begin{pmatrix}
			|\delta|^2 & \delta\gamma^*\sqrt{1-p} \\
			\delta^*\gamma\sqrt{1-p} & |\gamma|^2(1-p)
		\end{pmatrix}$ & $|\zeta_2^B\rangle$ & $|\eta_3^A\rangle$ \\  
		\hline
		$\rho_{23} = \frac{1}{(4-2p)^2} 
		\begin{pmatrix}
			|\alpha|^2 & -\alpha\beta^*\sqrt{1-p} \\
			-\alpha^*\beta\sqrt{1-p} & |\beta|^2(1-p)
		\end{pmatrix} \otimes \begin{pmatrix}
			|\delta|^2 & -\delta\gamma^*\sqrt{1-p} \\
			-\delta^*\gamma\sqrt{1-p} & |\gamma|^2(1-p)
		\end{pmatrix}$ & $|\zeta_2^B\rangle$ & $|\eta_4^A\rangle$ \\  
		\hline
			$\rho_{23} = \frac{1}{(4-2p)^2} 
		\begin{pmatrix}
			|\beta|^2 & \beta\alpha^*\sqrt{1-p} \\
			\beta^*\alpha\sqrt{1-p} & |\alpha|^2(1-p)
		\end{pmatrix} \otimes \begin{pmatrix}
		|\gamma|^2 & \gamma\delta^*\sqrt{1-p} \\
		\gamma^*\delta\sqrt{1-p} & |\delta|^2(1-p)
		\end{pmatrix}$ & $|\zeta_3^B\rangle$ & $|\eta_1^A\rangle$ \\ 
		\hline
			$\rho_{23} = \frac{1}{(4-2p)^2} 
		\begin{pmatrix}
			|\beta|^2 & \beta\alpha^*\sqrt{1-p} \\
			\beta^*\alpha\sqrt{1-p} & |\alpha|^2(1-p)
		\end{pmatrix} \otimes \begin{pmatrix}
			|\gamma|^2 & -\gamma\delta^*\sqrt{1-p} \\
			-\gamma^*\delta\sqrt{1-p} & |\delta|^2(1-p)
		\end{pmatrix}$ & $|\zeta_3^B\rangle$ & $|\eta_2^A\rangle$ \\ 
		\hline
			$\rho_{23} = \frac{1}{(4-2p)^2} 
		\begin{pmatrix}
			|\beta|^2 & \beta\alpha^*\sqrt{1-p} \\
			\beta^*\alpha\sqrt{1-p} & |\alpha|^2(1-p)
		\end{pmatrix}\otimes \begin{pmatrix}
		|\delta|^2 & -\delta\gamma^*\sqrt{1-p} \\
		-\delta^*\gamma\sqrt{1-p} & |\gamma|^2(1-p)
		\end{pmatrix}$ & $|\zeta_3^B\rangle$ & $|\eta_4^A\rangle$ \\  
		\hline
			$\rho_{23} = \frac{1}{(4-2p)^2} 
		\begin{pmatrix}
			|\beta|^2 & -\beta\alpha^*\sqrt{1-p} \\
			-\beta^*\alpha\sqrt{1-p} & |\alpha|^2(1-p)
		\end{pmatrix} \otimes \begin{pmatrix}
			|\gamma|^2 & \gamma\delta^*\sqrt{1-p} \\
			\gamma^*\delta\sqrt{1-p} & |\delta|^2(1-p)
		\end{pmatrix}$ & $|\zeta_4^B\rangle$ & $|\eta_1^A\rangle$ \\ 
		\hline
			$\rho_{23} = \frac{1}{(4-2p)^2} 
		\begin{pmatrix}
			|\beta|^2 & -\beta\alpha^*\sqrt{1-p} \\
			-\beta^*\alpha\sqrt{1-p} & |\alpha|^2(1-p)
		\end{pmatrix} \otimes \begin{pmatrix}
			|\gamma|^2 & -\gamma\delta^*\sqrt{1-p} \\
			-\gamma^*\delta\sqrt{1-p} & |\delta|^2(1-p)
		\end{pmatrix}$ & $|\zeta_4^B\rangle$ & $|\eta_2^A\rangle$ \\ 
		\hline
			$\rho_{23} = \frac{1}{(4-2p)^2} 
		\begin{pmatrix}
			|\beta|^2 & -\beta\alpha^*\sqrt{1-p} \\
			-\beta^*\alpha\sqrt{1-p} & |\alpha|^2(1-p)
		\end{pmatrix} \otimes \begin{pmatrix}
		|\delta|^2 & \delta\gamma^*\sqrt{1-p} \\
		\delta^*\gamma\sqrt{1-p} & |\gamma|^2(1-p)
		\end{pmatrix}$ & $|\zeta_4^B\rangle$ & $|\eta_3^A\rangle$ \\  
		\hline
	\end{tabular}
	\caption{Bob's and Alice's recovered non-normalized states ($\rho_{23}$) before weak-measurements, and the corresponding bell state measurement outcomes $|\zeta_j\rangle$ and $|\eta_i\rangle$ respectively}
	\label{tb1}
\end{table}
 Once Alice and Bob have obtained the recovered states ($\rho_{23}$) listed above in Tab.~\ref{tb1}, the two partners  invert the ADC effects by the application of weak measurements.  The weak measurement operators defined in \cite{eam3} are utilized to complete the optimized BT-EW scheme, however in our case we have separate weak measurement operators for the two partners. The complete set of weak measurement operators is $\{m_w^0, m_w^1 \}$, and is defined below 
\begin{eqnarray}
	m_w^0=\begin{bmatrix}
		\sqrt{1-p}&0\\
		0&1\\
	\end{bmatrix}~~~~~~ 	m_w^1=\begin{bmatrix}
	\sqrt{p}&0\\
	0&0\\
	\end{bmatrix}
\end{eqnarray}
Since Alice and Bob retain only the EAM outcomes corresponding to the invertible Kraus operator $k_0$, the weak measurement operator $m_w^0$ is applied to  reverse the ADC effect and enhance fidelity. The final teleported states are then normalized after the weak measurements.

In strategies utilizing weak measurements for decoherence free quantum teleportation a competition arises between success probability and fidelity of quantum teleportation. To achieve tunability of the weak measurement strength, we introduce a separate variable $q_w$, constrained by $ 0 \leq q_w \leq 1 $. Now, depending on the projective bell state measurement results, the two partners apply the unitary transformations equipped with weak measurement operators  in the last step of the BQT scheme. The combined operators are then defined as:
\begin{eqnarray}
M_i^A=U_i^Am_w^0=U_i^A\begin{pmatrix}
	\sqrt{1-q_w}&0\\
	0&1
\end{pmatrix},~~~~~M_j^B=U_j^Bm_w^0=U_j^B\begin{pmatrix}
\sqrt{1-q_w}&0\\
0&1
\end{pmatrix}~~~\quad i,j=1,2,3,4
\end{eqnarray}
where $U_i^A$ and $U_j^B$  are Alice's and Bob's unitary operators (pauli spin operators) acting on qubits $3$ and $2$ respectively, with $U_i^A, ~U_j^B= I (i,j=1), \sigma_z (i,j=2), \sigma_x (i,j=3)$ and $\sigma_x \sigma_z (i,j=4)$. Since the weak measurement operator $m_w^0$ acts independent of the bell state measurement results of the two partners, the operators $M_i^A $ and $M_J^B $ carry the dependence of these projective measurement results through $U_i^A$ and $U_j^B$ respectively. The final normalized teleported states are given by
\begin{eqnarray}
\rho_{M^{B}_{j}M^{A}_{i}}=\frac{M_i^AM_j^B\rho_{23}M_j^{B\dagger}M_i^{A\dagger}}{\text{Tr}(M_i^AM_j^B\rho_{23}M_j^{B\dagger}M_i^{A\dagger})}
\end{eqnarray}
where $\rho_{23}$ are the non-normalized states of the two partners defined in Eq.~\ref{nn} and listed in Tab.\ref{tb1}~ for sixteen possible projections. Moreover, the normalization factor $\text{Tr}(M_i^AM_j^B\rho_{23}M_j^{B\dagger}M_i^{A\dagger}) $ is the quantity of interest and is termed as the success probability of BQT, denoted by $ g_{M_j^BM_i^A}$. The sixteen possible normalized output states post weak measurements and unitary transformations are  
 \begin{eqnarray}
 	\rho_{M^{B}_j M^{A}_i} &=\frac{1}{(4-2p)^2g_{M_j^BM_i^A}}
 	\begin{cases}
 		\begin{array}{l}
 			\begin{pmatrix}
 				|\alpha|^2(1-q_
 	w) & \alpha\beta^*\sqrt{1-p}\sqrt{1-q_w} \\
 				\alpha^*\beta\sqrt{1-p}\sqrt{1-q_
 				w} & |\beta|^2(1-p)
 			\end{pmatrix}
 			\otimes
 			\begin{pmatrix}
 				|\gamma|^2(1-q_w) & \gamma\delta^*\sqrt{1-p}\sqrt{1-q_w} \\
 				\gamma^*\delta\sqrt{1-p}\sqrt{1-q_w} & |\delta|^2(1-p)
 			\end{pmatrix} \\[8pt] 
 			\hfill j,i=1,2
 		\end{array} \\[12pt]
 		
 		\begin{array}{l}
 			\begin{pmatrix}
 				|\alpha|^2(1-q_w) & \alpha\beta^*\sqrt{1-p}\sqrt{1-q_w} \\
 				\alpha^*\beta\sqrt{1-p}\sqrt{1-q_w} & |\beta|^2(1-p)
 			\end{pmatrix}
 			\otimes
 			\begin{pmatrix}
 				|\gamma|^2(1-p) & \gamma\delta^*\sqrt{1-p}\sqrt{1-q_w} \\
 				\gamma^*\delta\sqrt{1-p}\sqrt{1-q_w} & |\delta|^2(1-q_w)
 			\end{pmatrix} \\[8pt] 
 			\hfill j=1,2 \text{ and } i=3,4
 		\end{array} \\[12pt]
 		
 		\begin{array}{l}
 			\begin{pmatrix}
 				|\alpha|^2(1-p) & \alpha\beta^*\sqrt{1-p}\sqrt{1-q_w} \\
 				\alpha^*\beta\sqrt{1-p}\sqrt{1-q_w} & |\beta|^2(1-q_w)
 			\end{pmatrix}
 			\otimes
 			\begin{pmatrix}
 				|\gamma|^2(1-q_w) & \gamma\delta^*\sqrt{1-p}\sqrt{1-q_w} \\
 				\gamma^*\delta\sqrt{1-p}\sqrt{1-q_w} & |\delta|^2(1-p)
 			\end{pmatrix} \\[8pt] 
 			\hfill j=3,4 \text{ and } i=1,2
 		\end{array} \\[12pt]
 		
 		\begin{array}{l}
 			\begin{pmatrix}
 				|\alpha|^2(1-p) & \alpha\beta^*\sqrt{1-p}\sqrt{1-q_w} \\
 				\alpha^*\beta\sqrt{1-p}\sqrt{1-q_w} & |\beta|^2(1-q_w)
 			\end{pmatrix}
 			\otimes
 			\begin{pmatrix}
 				|\gamma|^2(1-p) & \gamma\delta^*\sqrt{1-p}\sqrt{1-q_w} \\
 				\gamma^*\delta\sqrt{1-p}\sqrt{1-q_w} & |\delta|^2(1-q_w)
 			\end{pmatrix} \\[8pt] 
 			\hfill j,i=3,4
 		\end{array}
 	\end{cases}
 	\label{out}
 \end{eqnarray} 
Clearly, if the strength of ADC is balanced by the weak measurement strength  i.e, ($q_w= p$) complete suppression of the noise occurs and the corresponding success probabilities of getting these states by Bob and Alice are  
\begin{eqnarray}
	g_{M^{B}_j M^{A}_i} &=&  
	\begin{cases}
		\begin{array}{l}
			\frac{1}{(4-2p)^2} \bigg[\bigg( |\alpha|^2(1-q_w) + |\beta|^2(1-p)\bigg) \bigg( |\gamma|^2(1-q_w) + |\delta|^2(1-p)\bigg) \bigg] \\[8pt] 
			\hfill j,i=1,2
		\end{array} \\[12pt]
		
		\begin{array}{l}
			\frac{1}{(4-2p)^2} \bigg[ \bigg(|\alpha|^2(1-q_w) + |\beta|^2(1-p)\bigg)  \bigg(|\gamma|^2(1-p) + |\delta|^2(1-q_w)\bigg) \bigg] \\[8pt] 
			\hfill j=1,2 \text{ and } i=3,4
		\end{array} \\[12pt]
		
		\begin{array}{l}
			\frac{1}{(4-2p)^2} \bigg[ \bigg(|\alpha|^2(1-p) + |\beta|^2(1-q_w)\bigg) \bigg( |\gamma|^2(1-q_w) + |\delta|^2(1-p) \bigg)\bigg] \\[8pt] 
			\hfill j=3,4 \text{ and } i=1,2
		\end{array} \\[12pt]
		
		\begin{array}{l}
			\frac{1}{(4-2p)^2} \bigg[ \bigg(|\alpha|^2(1-p) + |\beta|^2(1-q_w)\bigg)  \bigg(|\gamma|^2(1-p) + |\delta|^2(1-q_w) \bigg)\bigg] \\[8pt] 
			\hfill j,i=3,4
		\end{array}
	\end{cases}
\end{eqnarray}
which is the product of Bob's and Alice's success probabilities of recovering the input states. The total success probability of our protected BQT scheme  for complete suppression of ADC effects  is
\begin{eqnarray}
g_t^{BT-EW}=\sum_{j,i=1}^{4}	g_{M^{B}_j M^{A}_i}=\bigg(1-\frac{q_w}{2-p}\bigg)^2
\label{gt}
\end{eqnarray}
The superscript $ (\text{BT-EW}) $ used throughout this paper labels our protected BQT protocol aided with EAM and weak measurement. Clearly, the absence of weak measurement ($ q_w=0 $) implies the protocol is deterministic and occurs with $ g_t^{\text{BT-EW}}=1 $ (see Fig.~\ref{gt2d}).

Another quantity of interest which describes the performance of the BQT protocol is the fidelity of quantum teleportation measuring the overlap between the input and output states $F=\text{Tr}(\rho_{in}\rho_{out})$ . The fidelity between the output states of Bob and Alice Eq.~\ref{out} and each others input states Eq.~\ref{inp} is defined as 
\begin{eqnarray}
	F^{BA}_{ji} &=& \text{Tr} \bigg(\rho_a \otimes \rho_b \, \rho_{M^{B}_j M^{A}_i} \bigg) = F_j^B F_i^A, \quad \text{where}\nonumber \\
	F_j^B F_i^A &=&
	\begin{cases}
		\Bigg( \frac{|\beta|^4(1-p) + |\alpha|^4(1-q_w) + 2|\alpha|^2|\beta|^2\sqrt{1-q_w}\sqrt{1-p}}{|\alpha|^2(1-q_w) + |\beta|^2(1-p)} \Bigg)
		\Bigg( \frac{|\delta|^4(1-p) + |\gamma|^4(1-q_w) + 2|\gamma|^2|\delta|^2\sqrt{1-q_w}\sqrt{1-p}}{|\gamma|^2(1-q_w) + |\delta|^2(1-p)} \Bigg) & j,i=1,2 \\\\[10pt]
		
		\Bigg( \frac{|\beta|^4(1-p) + |\alpha|^4(1-q_w) + 2|\alpha|^2|\beta|^2\sqrt{1-q_w}\sqrt{1-p}}{|\alpha|^2(1-q_w) + |\beta|^2(1-p)} \Bigg)
		\Bigg( \frac{|\delta|^4(1-q_w) + |\gamma|^4(1-p) + 2|\gamma|^2|\delta|^2\sqrt{1-q_w}\sqrt{1-p}}{|\gamma|^2(1-p) + |\delta|^2(1-q_w)} \Bigg) &  j=1,2 \text{ and } i=3,4 \\\\[10pt]
		
		\Bigg( \frac{|\beta|^4(1-q_w) + |\alpha|^4(1-p) + 2|\alpha|^2|\beta|^2\sqrt{1-q_w}\sqrt{1-p}}{|\alpha|^2(1-p) + |\beta|^2(1-q_w)} \Bigg)
		\Bigg( \frac{|\delta|^4(1-p) + |\gamma|^4(1-q_w) + 2|\gamma|^2|\delta|^2\sqrt{1-q_w}\sqrt{1-p}}{|\gamma|^2(1-q_w) + |\delta|^2(1-p)} \Bigg) & j=3,4 \text{ and } i=1,2 \\\\[10pt]
		
		\Bigg( \frac{|\beta|^4(1-q_w) + |\alpha|^4(1-p) + 2|\alpha|^2|\beta|^2\sqrt{1-q_w}\sqrt{1-p}}{|\alpha|^2(1-p) + |\beta|^2(1-q_w)} \Bigg)
		\Bigg( \frac{|\delta|^4(1-q_w) + |\gamma|^4(1-p) + 2|\gamma|^2|\delta|^2\sqrt{1-q_w}\sqrt{1-p}}{|\gamma|^2(1-p) + |\delta|^2(1-q_w)} \Bigg)&  j,i=3,4
	\end{cases}
	\label{fid}
\end{eqnarray}
the total fidelity of BQT considering the probabilities of different projections Eq.~\ref{pro} is 
\begin{eqnarray}
\sum_{j,i=1}^{4}P_i^AP_j^BF_{ji}^{BA}
\label{tf}
\end{eqnarray}
Since the fidelity of BQT in Eq.~\ref{fid} depends on the input state parameters of the two partners, we define an input state independent quantity i.e, average fidelity to quantify the performance of our proposed BQT scheme  as
\begin{eqnarray}
F_{av}^{BT-EW}=\int_{0}^{1}\sum_{j,i=1}^{4}P_i^AP_j^BF_{ji}^{BA} d|\alpha|^2
\label{fav}
\end{eqnarray} 

For simplicity we have assumed that Alice and Bob's input states Eq.~\ref{inp} are arbitrary but the same, with $ \alpha=\gamma$ and $ \beta=\delta$ . The behaviour of $F_{av}^{BT-EW}$ in Eq.(\ref{fav}) is shown in Fig.~\ref{fiall} and reflects an interplay between strength of ADC and  weak measurement. In order, to analyze the suppression of ADC effects by weak measurements in the final step of the BQT scheme, we consider the Eqs.~\ref{gt}~-~\ref{fav}. It is observed that complete suppression of ADC effects occurs for $q_w=p$, resulting in perfect BQT marked by $F_{av}^{BT-EW}=1$, and the corresponding teleportation success probability $ g_t^{BT-EW}=\big(1-\frac{p}{2-p}\big)^2$. Also, in absence of  weak measurements i.e  $ q_w=0$, BQT occurs with unit success probability at the price of  reduced fidelity. By adjusting the weak measurement strength within the range $q_w \in (0, p)$, a balance between $F_{av}^{BT-EW}$ and $g_{t}^{BT-EW}$ can be achieved. Additionally, we have determined that the values of weak measurement strength $q_w \in (p, 1]$ are strictly prohibited, as both $F_{av}^{BT-EW}$ and $g_{t}^{BT-EW}$ are lower than at $q_w = p$ (see Figs.~\ref{fiall} and \ref{gall}).
\subsubsection{Numerical Results and Analysis}
\label{ssub1}
\begin{figure}[H]
	\centering
	\begin{subfigure}[b]{0.45\textwidth}
		\includegraphics[width=9cm, height=7cm]{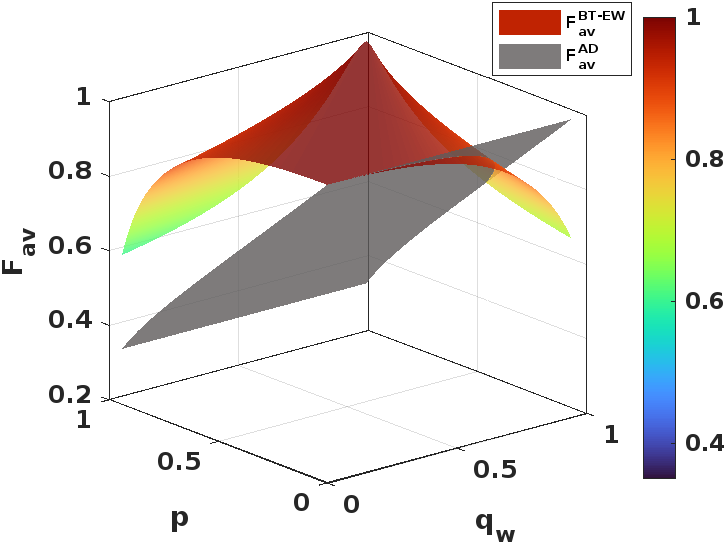}
		\caption{}
		\label{F3d}
	\end{subfigure}
	\hspace{0.05\textwidth}
	\begin{subfigure}[b]{0.45\textwidth}
		\includegraphics[width=\textwidth, height=7cm]{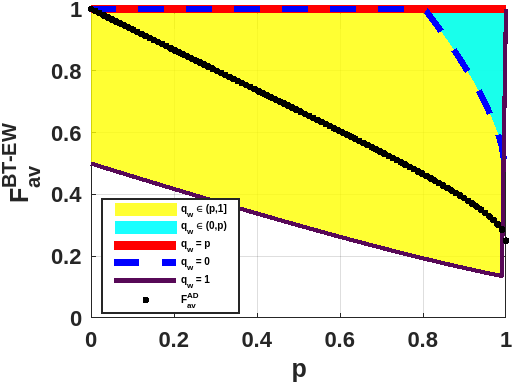}
		\caption{}
		\label{F2d}
	\end{subfigure}
	\caption{(\ref{F3d})  Average fidelity $F_{av} $ as a function of decoherence rate $p$ and the weak measurement strength $q_w$ when noise affects only recovery qubits. The red plane shows average fidelity $F_{av}^{BT-EW}$, when ADC effects are neutralized by EAM and weak measurements. The gray plane is the average fidelity $F_{av}^{AD}$ of BQT in absence of protection.  (\ref{F2d}) Comparison of average teleportation fidelities as a function of $ p$ at different values of $ q_w$. The cyan shaded region depicts the average fidelity for $q_w\in (0,p) $ and shows the perfect balance between noise and protection, the yellow shaded region shows prohibited region  $q_w\in(p,1]$, with $q_w=1$ shown by solid purple line marks its boundary. The red solid line shows perfect teleportation occurs when $ q_w=p$ with unit fidelity.  The black dotted line is average fidelity in absence of EAM and weak measurement lying below $q_w=0$ blue dashed line, implying EAM is effective in channel protection.}
	\label{fiall}
\end{figure}

\begin{figure}[H]
	\centering
	\begin{subfigure}[b]{0.45\textwidth}
		\includegraphics[width=9cm, height=6cm]{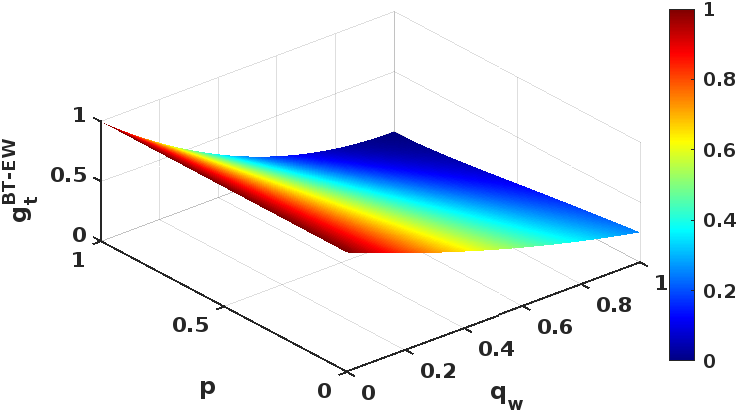}
		\caption{}
		\label{gt3d}
	\end{subfigure}
	\hspace{0.05\textwidth}
	\begin{subfigure}[b]{0.45\textwidth}
		\includegraphics[width=8cm, height=6cm]{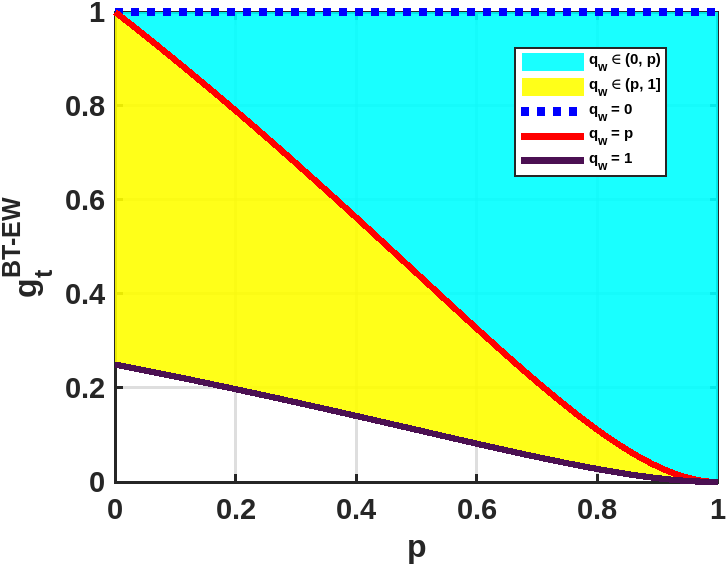}
		\caption{}
		\label{gt2d}
	\end{subfigure}
	\caption{(\ref{gt3d}) Total teleportation success probability of proposed BQT protocol as a function of the weak measurement strength $q_w$ and decay rate $p$. (\ref{gt2d} ) Comparison of total teleportation success probability  with respect to $p$,the cyan shaded region corresponds to $g_t^{BT-EW}$ for $ q_w\in(0,p) $, the yellow shaded region is the prohibited region $q_w\in(p,1]$ with boundary $q_w=1$ (purple line), blue dashed line $q_w=0$ and red solid line $q_w=p$.}
	\label{gall}
\end{figure}

The key features of our protected BQT scheme are discussed below. 
For comparative analysis, we consider the average fidelity of BQT through ADC without EAM and weak measurements as shown by the gray plane in Fig.~\ref{F3d}, and is given by.
\begin{eqnarray}
F^{AD}_{av}=\frac{1}{9}\bigg(2-\frac{p}{2}+\sqrt{1-p}\bigg)^2
	\label{unp}
\end{eqnarray}
The BQT protocol without protection through ADC operates deterministically due to the absence of weak measurements, as discussed in Appendix~\ref{apB}. In 	Fig.~\ref{fiall}, average fidelity of our protected BQT protocol Eq.~\ref{fav}, is plotted with Fig.~\ref{F3d} depicting comparison of $F_{av}^{BT-EW}$ and that of unprotected BQT scheme $F_{av}^{AD}$ Eq.~\ref{unp}.  The behaviour is analyzed by tuning the strength of ADC and weak measurements. Also Fig.~\ref{F2d}, shows the optimization of average fidelity for different values of weak measurement strength against ADC effects. In addition, the behaviour of teleportation success probability Eq.~\ref{gt}  is plotted in Fig.~\ref{gall} as a function of the damping strength $p$ and the weak measurement strength $q_w$. Since unprotected BQT occurs deterministically we have only shown teleportation success probability of our BQT protocol in Fig.~\ref{gt3d} and its optimization for different $q_w$ values with respect to decay rate $p$ in Fig.~\ref{gt2d}.

In Fig.~\ref{F3d}, a remarkable enhancement in average fidelity $F_{av}^{BT-EW}$ depicted by red plane is observed over unprotected BQT scheme (gray plane) for all decaying rates $p$, signifying the effectiveness of  protection. By analyzing Fig.~\ref{F2d}, we establish the optimization of average fidelity by comparing it at different $q_w$ values. The black dotted line represents the average fidelity $F_{av}^{AD}$ in the absence of protection. The blue dashed line at $q_w=0$ signifies the effectiveness of EAM in channel protection. The red solid line at $q_w=p$ marks the point of optimum average fidelity, $F_{av}^{BT-EW}=1$. The cyan shaded region, corresponding to $q_w \in (0,p)$, depicts the balance between noise and protection. The yellow shaded region represents the prohibited region, $q_w \in (p,1]$, with the boundary $q_w=1$ indicated by the purple line, since $F_{av}^{BT-EW}$ is less than the optimum value in this range. It is clear that the average fidelity is remarkably enhanced by protection. Furthermore, even in the absence of weak measurements ($q_w=0$), higher average fidelity is observed, implying that EAMs are effective in protecting the channel against ADC effects.
 Perfect BQT is achieved for $q_w=p$, with $F_{av}^{BT-EW}=1$ corresponds to complete suppression of ADC effects and marks the optimization of the BQT scheme. Also, it is observed when weak measurement strength $q_w$, exceeds ADC strength $p$, a lower value of average fidelity is observed than at $q_w=p$, marking the prohibited domain $q_w\in(p,1]$.

%\vspace{6cm}
Moreover, in Fig.~\ref{gt3d},total success probability of BQT is observed to remain non zero across a wide range of ADC and weak measurement strengths. It is clear that $g_t^{BT-EW}$ undergoes a smooth decay as $q_w$ is increased and exhibits a reverse trend as that of $F_{av}^{BT-EW}$. Also Fig.~\ref{gt2d}, shows the behaviour of $g_{t}^{BT-EW}$ for the same $q_w$ values as for $F_{av}^{BT-EW}$, with blue dashed line $q_w=0$ marks $g_{t}^{BT-EW}=1$ shows BQT is achieved deterministically with very high fidelity, red solid line $q_w=p$ marks optimization of BQT scheme. Again it is clear from yellow shaded region  $q_w\in(p,1]$  that $g_t^{BT-EW}$ is lower than at $q_w=p$ and marks prohibited domain of $q_w$.  It is clear that a decrease in $g_{t}^{BT-EW}$ is observed when $q_w$ increases, however a corresponding reverse trend is observed for $F_{av}^{BT-EW}$.  Since the weak measurement can be controlled externally and by setting $q_w~\in(0,p)$ a balance between  $F_{av}^{BT-EW}$ and  $g_{t}^{BT-EW}$ is achieved depicted by cyan shaded region. Thereby,  the optimization strategy of the proposed protocol involves tuning the weak measurement strength within this range to achieve high fidelity along with a non-zero teleportation success probability.
\section{BQT with Amplitude Damping on all Qubits}
\label{sub2}
\begin{figure}[H]
	\centering
	\tikzset{
		noisy/.style={
			starburst,
			fill=lime,
			draw=blue,
			line width=.5pt, % Make the starburst border thicker
			inner xsep=-4pt,
			inner ysep=-1pt
		},
		quantum wire/.style={thick}, % Make wires thicker
		gate/.style={draw, thick, minimum size=3em}, % Make gates bold
		label style/.style={font=\bfseries\boldmath} % Bold labels
	}
	\begin{quantikz}[row sep=4mm, column sep=0.5mm]
		\lstick{$\rho_{a}$}&\qw&\qw&\qw&\qw&\qw&\qw&\qw&\qw&\qw&\qw&\qw&\qw&\qw&\qw&\qw&\qw&\qw&\qw&\qw&\qw&\qw&\qw&\qw&\qw&\qw&\qw&\qw&\qw&\qw&\qw&\qw&\qw&\qw&\qw&\qw&\qw&\qw&\qw&\qw&\qw&\qw&\qw&\qw&\meter[style={fill=yellow!60, draw=orange, thick}]{}-
		&\setwiretype{c}&&&&&&&&&&\ctrl[vertical
		wire=c]{5}\\\\\\\\
		\lstick{$|0\rangle_1$}&\qw&\qw &\qw&\qw&\gate[style={draw=black, fill=green!20}]{H}&\qw&\qw&\qw&\qw&\qw&\qw&\qw&\ctrl{1}&\qw&\qw&\qw&\qw&\qw&\qw&\qw&\qw&\qw&\qw&\qw&\qw&\qw&\gate[1,style={noisy},label
		style=black]{\text{ADC}}&\qw&\qw&\qw&\qw&\qw&\qw&\qw&\gate[1,style={fill=pink!80, draw=blue, label style={text=red}}]{\text{EAM}}&\qw&\qw&\qw&\qw&\qw&\qw&\qw&\qw&\meter[style={fill=yellow!60, draw=orange, thick}]{}-
		&\setwiretype{c}&&&&&&&&&&\control\cw\\
		\lstick{$|0\rangle_2$}&\qw&\qw&\qw&\qw&\qw&\qw&\qw&\qw&\qw&\qw&\qw&\qw&\targ{}&\qw&\qw&\qw&\qw&\qw&\qw&\qw&\qw&\qw&\qw&\qw&\qw&\qw&\gate[1,style={noisy},label
		style=black]{\text{ADC}}&\qw&\qw&\qw&\qw&\qw&\qw&\qw&\gate[1,style={fill=pink!80, draw=blue, label style={text=red}}]{\text{EAM}}&\qw&\qw&\qw&\qw&\qw&\qw&\qw&\qw&\qw&\qw&\qw&\qw&\qw&\qw&\qw&\qw&\qw&\qw&\qw&\gate[1,style={fill=lime!80, draw=blue, label style={text=red}}]{\text{$U_j^B m_{w}^{0'}$}}&\qw&\qw&\qw&\qw&\qw
		&\qw&\qw&\qw&\qw&\qw&\qw&\qw&\qw&\qw&\qw&\qw&\qw&\qw&\qw&\qw&\qw&\qw&\qw&\qw&\qw&\qw&~~~ \rstick[2]{\textcolor{red}{$~~~\rho'_{M_j^B M_i^A}$}}   
		\\
		\lstick{$|0\rangle_3$}&\qw&\qw &\qw&\qw&\gate[style={draw=black, fill=green!20}]{H}&\qw&\qw&\qw&\qw&\qw&\qw&\qw&\ctrl{1}&\qw&\qw&\qw&\qw&\qw&\qw&\qw&\qw&\qw&\qw&\qw&\qw&\qw&\gate[1,style={noisy},label
		style=black]{\text{ADC}}&\qw&\qw&\qw&\qw&\qw&\qw&\qw&\gate[1,style={fill=pink!80, draw=blue, label style={text=red}}]{\text{EAM}}&\qw&\qw&\qw&\qw&\qw&\qw&\qw&\qw&\qw&\qw&\qw&\qw&\qw&\qw&\qw&\qw&\qw&\qw&\qw&\gate[1,style={fill=lime!80, draw=blue, label style={text=red}}]{\text{$U_i^A m_{w}^{0'}$}}
		&\qw&\qw&\qw&\qw&\qw
		&\qw&\qw&\qw&\qw&\qw&\qw&\qw&\qw&\qw&\qw&\qw&\qw&\qw&\qw&\qw&\qw&\qw&\qw&\qw&\qw&\qw&\qw~~\\
		\lstick{$|0\rangle_4$}&\qw&\qw&\qw&\qw&\qw&\qw&\qw&\qw&\qw&\qw&\qw&\qw&\targ{}&\qw&\qw&\qw&\qw&\qw&\qw&\qw&\qw&\qw&\qw&\qw&\qw&\qw&\gate[1,style={noisy},label
		style=black]{\text{ADC}}&\qw&\qw&\qw&\qw&\qw&\qw&\qw&\gate[1,style={fill=pink!80, draw=blue, label style={text=red}}]{\text{EAM}}&\qw&\qw&\qw&\qw&\qw&\qw&\qw&\qw&\meter[style={fill=yellow!60, draw=orange, thick}]{}-
		&\setwiretype{c}&&&&&&&&&&\control\cw\\\\\\\\
		\lstick{$\rho_{b}$}&\qw&\qw&\qw&\qw&\qw&\qw&\qw&\qw&\qw&\qw&\qw&\qw&\qw&\qw&\qw&\qw&\qw&\qw&\qw&\qw&\qw&\qw&\qw&\qw&\qw&\qw&\qw&\qw&\qw&\qw&\qw&\qw&\qw&\qw&\qw&\qw&\qw&\qw&\qw&\qw&\qw&\qw&\qw&\meter[style={fill=yellow!60, draw=orange, thick}]{}-
		&\setwiretype{c}&&&&&&&&&&\ctrl[vertical
		wire=c]{-5}\\	
	\end{quantikz}
	\begin{tikzpicture}[overlay]
		\node[draw=black,very thick, fit={(-11.8,3.5) (-11.2,4.4)}, inner sep=1mm, label=above:{ Alice's input state}] {};
		\node[draw=blue,very thick, fit={(-11.8,2.4) (-9.5,-1.6)}, inner sep=1mm, label=below:{(a) Preparation}, label=above:{ $| \phi\rangle_{1234}$}] {};
		\node[draw=red,very thick,dotted, fit={(-9.2,-2) (-6.3,2.6)}, inner sep=1mm,label=below:{(b) Distribution} ,label=above:{ $ \rho^{f'}_{1234}$}] {};
		\node[draw=cyan, very thick, dashed, dash pattern=on 5pt off 5pt, fit={(0.1,-3.8) (-6,4.4)}, inner sep=1mm,label=above:{(c)Protected BQT}] {};
		\node[draw=black,very thick, fit={(-11.8,-2.9) (-11.2,-3.7)}, inner sep=1mm, label=below:{ Bob's input state}] {};
	\end{tikzpicture}
	\caption{ Quantum circuit of protected BQT when all  the four channel qubits are passed through ADCs and subjected to EAMs. In the circuit (a) Four qubit channel preparation (b) Entanglement distribution by Charlie  (c) Protected BQT with weak measurements. The modified unitary operators equipped with weak measurements  $U_i^A m_w^{0'}$ and $U_j^B m_w^{0'}$  are  applied on recovery qubits ($3$ and $2$) respectively. The teleported state post protection is $\rho'_{M^{B}_j M^{A}_i}$.}
	\label{ebqt}
\end{figure}
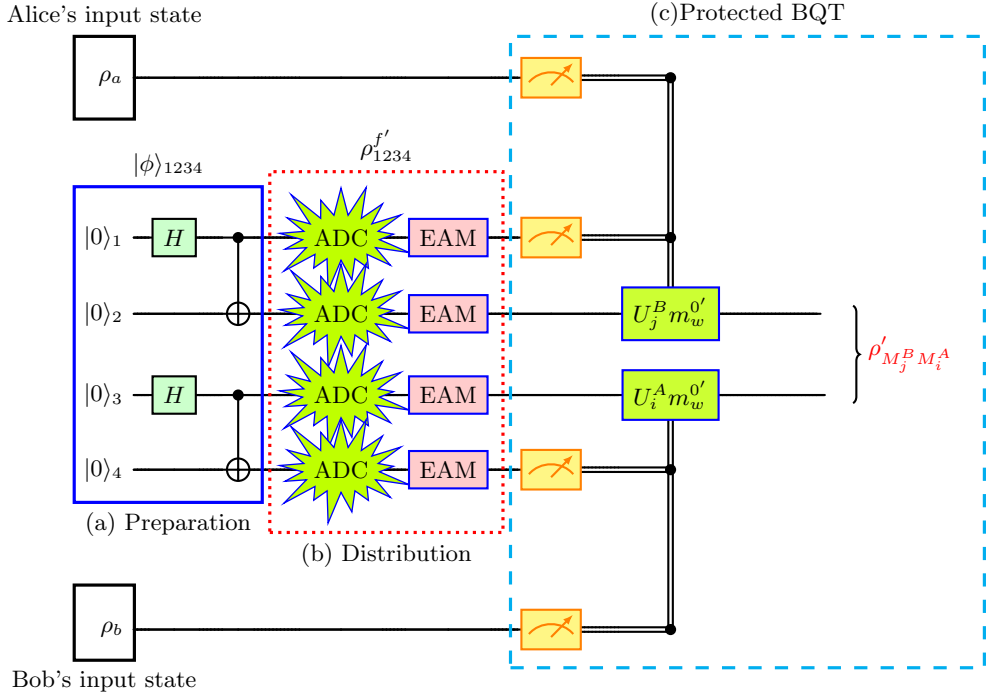

In this section, the BQT scheme is analyzed when the entire four-qubit channel, given by Eq.~\ref{eqd}, is subjected to noise. The protocol incorporates EAM and weak measurements, following the same principles as discussed in Sec.~\ref{sub1}.  The quantum circuit of the proposed protocol is shown in Fig.~\ref{ebqt}. Here the third party sends all four qubits through independent ADCs having same decay rate $p$. Again charlie, sends qubits marked $(1,3)$ to Alice and $(2,4)$ to Bob as in Sec.~\ref{sub1}. Both partners apply EAMs to their respective channel qubits. After the entangled channel is established between Alice and Bob, the BQT protocol is initiated and completed with the application of weak measurements in the final step. The proposed protocol is discussed below

 Since, all the four qubits are exposed to decoherence, the effect is portrayed through four qubit Kraus operators defined separately for Alice's and Bob's qubits respectively as

\begin{eqnarray}
	K_{ii'}^A = k_{i} \otimes I \otimes k_{i'} \otimes I,~~~~~ 
	K_{jj'}^B = I \otimes k_j \otimes I \otimes k_{j'} , \quad (i,i'\& j,j' = 0,1).
\end{eqnarray}
With $k_{i,i'}$ and $k_{j,j'}$ being single-qubit Kraus operators given by Eq.~\ref{ko}. Once Charlie transmits the respective qubits to Alice and Bob, they employ EAMs and communicate the measurement outcomes corresponding to the quantum trajectory with the invertible Kraus operators $K_{00}^A$ and $K_{00}^B$. This process leads to the formation of the four-qubit entangled channel between the two partners and is given by.
\begin{eqnarray}
	\rho_{1234}^{f'} &=& \frac{K_{00}^A K_{00}^B \rho_{1234} K_{00}^{B\dagger} K_{00}^{A\dagger}}{\text{Tr}(K_{00}^A K_{00}^B \rho_{1234} K_{00}^{B\dagger} K_{00}^{A\dagger})} 
\end{eqnarray}
	\begin{equation}
	%\footnotesize
	\rho_{\text{1234}}^{f'} =\frac{1}{4g_{1234}^{EAM}} \begin{pmatrix}
		1 & \cdots & (1-p) & \cdots &(1-p)& \cdots & (1-p)^2 \\
		\vdots & \cdots & \vdots & \cdots & \vdots & \cdots & \vdots \\
		(1-p) & \cdots & (1-p)^2 & \cdots & (1-p)^2 & \cdots & (1-p)^3\\
		\vdots & \cdots & \vdots & \cdots & \vdots & \cdots & \vdots \\
		\vdots & \cdots & \vdots & \cdots & \vdots & \cdots & \vdots \\
		(1-p)& \cdots & (1-p)^2 & \cdots & (1-p)^2& \cdots & (1-p)^3 \\
		\vdots & \cdots & \vdots & \cdots & \vdots & \cdots & \vdots \\
		(1-p)^2 & \cdots & (1-p)^3 & \cdots & (1-p)^3 & \cdots & (1-p)^4
	\end{pmatrix}
	\label{wchem}
\end{equation}
where, rest of the elements being zero are not shown and  $g_{1234}^{EAM}$ quantifies the entanglement distribution success probability in the symmetric exposure of four qubit quantum channel to ADC effects and is given as
\begin{eqnarray}
	g_{1234}^{EAM}=\text{Tr}(K_{00}^A K_{00}^B \rho_{1234} K_{00}^{B\dagger} K_{00}^{A\dagger})=\frac{\big((1-p)^2+1\big)^2}{4}
\end{eqnarray}
Once the four-qubit quantum channel given in Eq.~\ref{wchem} is successfully distributed and established between the two partners, the BQT protocol proceeds following the same steps as discussed in Sec.~\ref{sub1}. The only distinction arises in the final step, where the weak measurement operators are specifically designed to counteract the effects of ADC.

Now, the joint probability of measuring the bell states Eq.~\ref{bst}, by Alice and Bob is 
\begin{eqnarray}
	\text{P}_{i}^{A'} \text{P}_{j}^{B'} &=\frac{1}{4\big(1+(1-p)^2\big)^2}& 
	\left\{
	\begin{aligned}
		& \bigg( |\beta|^2(p^2-2p)+1\bigg)\times \bigg(|\delta|^2(p^2-2p)+1 \bigg) \quad &\text{for } i,j=1,2 \\
	& \bigg( |\beta|^2(p^2-2p)+1\bigg)\times \bigg(|\gamma|^2(p^2-2p)+1 \bigg) \quad &\text{for } i=1,2 \text{ \& } j=3,4 \\
	& \bigg( |\alpha|^2(p^2-2p)+1\bigg)\times \bigg(|\delta|^2(p^2-2p)+1 \bigg) \quad &\text{for } i=3,4 \text{ \& } j=1,2 \\
	& \bigg( |\alpha|^2(p^2-2p)+1\bigg)\times \bigg(|\gamma|^2(p^2-2p)+1 \bigg)\quad &\text{for } i,j=3,4
	\end{aligned}
	\right.
	\label{epro}
\end{eqnarray}
by classically communicating each other the projective measurement outcomes, the recovered states of the two partners are obtained following the operations defined in Eq.~\ref{nn}, and are listed in Tab.~\ref{tb2} along with corresponding measurement outcomes.

\begin{table}[H]
	\centering
	\begin{tabular}{|c|c|c|}
		\hline
		\multirow{2}{*}{Bob's and Alice's Recovered state ($\rho^{'}_{23}$)} & \multicolumn{2}{c|}{Measurement Outcomes} \\ 
		\cline{2-3}  
		&~~~~~~~~~~ Bob & Alice \\  
	%	\hline
		(Noise on all qubits)& & \\  
		\hline
		$\rho^{'}_{23} = \frac{1}{4\big(1+(1-p)^2\big)^2} 
		\begin{pmatrix}
			|\alpha|^2 & \alpha\beta^*(1-p) \\
			\alpha^*\beta(1-p) & |\beta|^2(1-p)^2
		\end{pmatrix} \otimes \begin{pmatrix}
			|\gamma|^2 & \gamma\delta^*(1-p) \\
			\gamma^*\delta(1-p) & |\delta|^2(1-p)^2
		\end{pmatrix}$ & $|\zeta_1^B\rangle$ & $|\eta_1^A\rangle$ \\ 
		\hline
			$\rho^{'}_{23} = \frac{1}{4\big(1+(1-p)^2\big)^2} 
		\begin{pmatrix}
			|\alpha|^2 & -\alpha\beta^*(1-p) \\
			-\alpha^*\beta(1-p) & |\beta|^2(1-p)^2
		\end{pmatrix} \otimes \begin{pmatrix}
			|\gamma|^2 & -\gamma\delta^*(1-p) \\
			-\gamma^*\delta(1-p) & |\delta|^2(1-p)^2
		\end{pmatrix}$  & $|\zeta_2^B\rangle$ & $|\eta_2^A\rangle$ \\ 
		\hline
		$\rho^{'}_{23} = \frac{1}{4\big(1+(1-p)^2\big)^2} 
		\begin{pmatrix}
			|\beta|^2 & \beta\alpha^*(1-p) \\
			\beta^*\alpha(1-p) & |\alpha|^2(1-p)^2
		\end{pmatrix} \otimes \begin{pmatrix}
			|\delta|^2 & \delta\gamma^*(1-p) \\
			\delta^*\gamma(1-p) & |\gamma|^2(1-p)^2
		\end{pmatrix}$ & $|\zeta_3^B\rangle$ & $|\eta_3^A\rangle$ \\ 
		\hline
		$\rho^{'}_{23} = \frac{1}{4\big(1+(1-p)^2\big)^2} 
	\begin{pmatrix}
		|\beta|^2 & -\beta\alpha^*(1-p) \\
		-\beta^*\alpha(1-p) & |\alpha|^2(1-p)^2
	\end{pmatrix} \otimes \begin{pmatrix}
		|\delta|^2 & -\delta\gamma^*(1-p) \\
		-\delta^*\gamma(1-p) & |\gamma|^2(1-p)^2
	\end{pmatrix}$ & $|\zeta_4^B\rangle$ & $|\eta_4^A\rangle$ \\ 
		\hline
	$\rho^{'}_{23} = \frac{1}{4\big(1+(1-p)^2\big)^2} 
	\begin{pmatrix}
		|\alpha|^2 & \alpha\beta^*(1-p) \\
		\alpha^*\beta(1-p) & |\beta|^2(1-p)^2
	\end{pmatrix} \otimes \begin{pmatrix}
		|\gamma|^2 & -\gamma\delta^*(1-p) \\
		-\gamma^*\delta(1-p) & |\delta|^2(1-p)^2
	\end{pmatrix}$ & $|\zeta_1^B\rangle$ & $|\eta_2^A\rangle$ \\ 
		\hline
	$\rho^{'}_{23} = \frac{1}{4\big(1+(1-p)^2\big)^2} 
	\begin{pmatrix}
		|\alpha|^2 & -\alpha\beta^*(1-p) \\
		-\alpha^*\beta(1-p) & |\beta|^2(1-p)^2
	\end{pmatrix} \otimes \begin{pmatrix}
		|\gamma|^2 & \gamma\delta^*(1-p) \\
		\gamma^*\delta(1-p) & |\delta|^2(1-p)^2
	\end{pmatrix}$ & $|\zeta_2^B\rangle$ & $|\eta_1^A\rangle$ \\ 
		\hline
			$\rho^{'}_{23} = \frac{1}{4\big(1+(1-p)^2\big)^2} 
		\begin{pmatrix}
			|\alpha|^2 & \alpha\beta^*(1-p) \\
			\alpha^*\beta(1-p) & |\beta|^2(1-p)^2
		\end{pmatrix} \otimes \begin{pmatrix}
		|\delta|^2 & \delta\gamma^*(1-p) \\
		\delta^*\gamma(1-p) & |\gamma|^2(1-p)^2
		\end{pmatrix}$  & $|\zeta_1^B\rangle$ & $|\eta_3^A\rangle$ \\  
		\hline
		$\rho^{'}_{23} = \frac{1}{4\big(1+(1-p)^2\big)^2} 
	\begin{pmatrix}
		|\alpha|^2 & \alpha\beta^*(1-p) \\
		\alpha^*\beta(1-p) & |\beta|^2(1-p)^2
	\end{pmatrix} \otimes \begin{pmatrix}
		|\delta|^2 & -\delta\gamma^*(1-p) \\
		-\delta^*\gamma(1-p) & |\gamma|^2(1-p)^2
	\end{pmatrix}$  & $|\zeta_1^B\rangle$ & $|\eta_4^A\rangle$ \\  
		\hline
		$\rho^{'}_{23} = \frac{1}{4\big(1+(1-p)^2\big)^2} 
	\begin{pmatrix}
		|\alpha|^2 & -\alpha\beta^*(1-p) \\
	-\alpha^*\beta(1-p) & |\beta|^2(1-p)^2
	\end{pmatrix} \otimes \begin{pmatrix}
		|\delta|^2 & \delta\gamma^*(1-p) \\
		\delta^*\gamma(1-p) & |\gamma|^2(1-p)^2
	\end{pmatrix}$  & $|\zeta_2^B\rangle$ & $|\eta_3^A\rangle$ \\  
		\hline
		$\rho^{'}_{23} = \frac{1}{4\big(1+(1-p)^2\big)^2} 
	\begin{pmatrix}
		|\alpha|^2 & -\alpha\beta^*(1-p) \\
		-\alpha^*\beta(1-p) & |\beta|^2(1-p)^2
	\end{pmatrix} \otimes \begin{pmatrix}
		|\delta|^2 & -\delta\gamma^*(1-p) \\
		-\delta^*\gamma(1-p) & |\gamma|^2(1-p)^2
	\end{pmatrix}$  & $|\zeta_2^B\rangle$ & $|\eta_4^A\rangle$ \\  
		\hline
		$\rho^{'}_{23} = \frac{1}{4\big(1+(1-p)^2\big)^2} 
	\begin{pmatrix}
		|\beta|^2 & \beta\alpha^*(1-p) \\
		\beta^*\alpha(1-p) & |\alpha|^2(1-p)^2
	\end{pmatrix} \otimes \begin{pmatrix}
	|\gamma|^2 & \gamma\delta^*(1-p) \\
	\gamma^*\delta(1-p) & |\delta|^2(1-p)^2
	\end{pmatrix}$& $|\zeta_3^B\rangle$ & $|\eta_1^A\rangle$ \\ 
		\hline
		$\rho^{'}_{23} = \frac{1}{4\big(1+(1-p)^2\big)^2} 
	\begin{pmatrix}
		|\beta|^2 & \beta\alpha^*(1-p) \\
		\beta^*\alpha(1-p) & |\alpha|^2(1-p)^2
	\end{pmatrix} \otimes \begin{pmatrix}
		|\gamma|^2 & -\gamma\delta^*(1-p) \\
		-\gamma^*\delta(1-p) & |\delta|^2(1-p)^2
	\end{pmatrix}$ & $|\zeta_3^B\rangle$ & $|\eta_2^A\rangle$ \\ 
		\hline
			$\rho^{'}_{23} = \frac{1}{4\big(1+(1-p)^2\big)^2} 
		\begin{pmatrix}
			|\beta|^2 & \beta\alpha^*(1-p) \\
			\beta^*\alpha(1-p) & |\alpha|^2(1-p)^2
		\end{pmatrix} \otimes \begin{pmatrix}
		|\delta|^2 & -\delta\gamma^*(1-p) \\
		-\delta^*\gamma(1-p) & |\gamma|^2(1-p)^2
		\end{pmatrix}$  & $|\zeta_3^B\rangle$ & $|\eta_4^A\rangle$ \\  
		\hline
		$\rho^{'}_{23} = \frac{1}{4\big(1+(1-p)^2\big)^2} 
	\begin{pmatrix}
		|\beta|^2 & -\beta\alpha^*(1-p) \\
		-\beta^*\alpha(1-p) & |\alpha|^2(1-p)^2
	\end{pmatrix} \otimes\begin{pmatrix}
	|\gamma|^2 & \gamma\delta^*(1-p) \\
	\gamma^*\delta(1-p) & |\delta|^2(1-p)^2
	\end{pmatrix}$ & $|\zeta_4^B\rangle$ & $|\eta_1^A\rangle$ \\ 
		\hline
			$\rho^{'}_{23} = \frac{1}{4\big(1+(1-p)^2\big)^2} 
		\begin{pmatrix}
			|\beta|^2 & -\beta\alpha^*(1-p) \\
			-\beta^*\alpha(1-p) & |\alpha|^2(1-p)^2
		\end{pmatrix} \otimes\begin{pmatrix}
			|\gamma|^2 & -\gamma\delta^*(1-p) \\
			-\gamma^*\delta(1-p) & |\delta|^2(1-p)^2
		\end{pmatrix}$ & $|\zeta_4^B\rangle$ & $|\eta_2^A\rangle$ \\ 
		\hline
		$\rho^{'}_{23} = \frac{1}{4\big(1+(1-p)^2\big)^2} 
	\begin{pmatrix}
		|\beta|^2 & -\beta\alpha^*(1-p) \\
		-\beta^*\alpha(1-p) & |\alpha|^2(1-p)^2
	\end{pmatrix} \otimes \begin{pmatrix}
		|\delta|^2 & \delta\gamma^*(1-p) \\
		\delta^*\gamma(1-p) & |\gamma|^2(1-p)^2
	\end{pmatrix}$ & $|\zeta_4^B\rangle$ & $|\eta_3^A\rangle$ \\  
		\hline
	\end{tabular}
\caption{Non-normalized recovered states ($\rho^{'}_{23}$) of Bob and Alice when the entire four-qubit channel is exposed to noise, prior to weak measurements, along with the corresponding Bell state measurement outcomes $|\zeta_j\rangle$ and $|\eta_i\rangle$.}
	\label{tb2}
\end{table}
\vspace{0.2cm}
To successfully complete the BQT protocol and ensure that Alice and Bob recover the optimized teleported states, weak measurement operators are carefully designed to counteract the effects of ADC in the final step as. 
\begin{eqnarray}
	M_i^{A'}=U_i^A m_w^{0'}=U_i^A\begin{pmatrix}
		(1-q_w)&0\\
		0&1
	\end{pmatrix},~~~~~M_j^{B'}=U_j^B m_w^{0'}=U_j^B\begin{pmatrix}
		(1-q_w)&0\\
		0&1
	\end{pmatrix}~~~\quad i,j=1,2,3,4
	\label{ewm}
\end{eqnarray}
where $U_i^A$ and  $U_j^B$ are the Pauli spin operators already defined in Sec.~\ref{sub1} and $q_w\in[0,1]$, is the weak measurement strength. Finally, Bob and Alice apply weak measurements equipped with unitary transformations Eq.~\ref{ewm} on the recovered states given in Tab.~\ref{tb2} and the optimized teleported states of the two partners takes the form
\begin{eqnarray}
	\rho^{'}_{M^{B}_j M^{A}_i} &= \frac{1}{4\big(1+(1-p)^2\big)^2g^{'}_{M_j^B M_i^A}} &
	\begin{cases}
		\begin{array}{l}
			\begin{pmatrix}
				|\alpha|^2(1-q_w)^2 & \alpha\beta^*(1-p)(1-q_w) \\
				\alpha^*\beta(1-p)(1-q_w)  & |\beta|^2(1-p)^2
			\end{pmatrix} 
			\otimes
			\begin{pmatrix}
				|\gamma|^2(1-q_w) & \gamma\delta^*(1-p)(1-q_w)  \\
				\gamma^*\delta(1-p)(1-q_w) & |\delta|^2(1-p)^2
			\end{pmatrix}\\[8pt]
			 \hfill j,i=1,2
		\end{array} \\[12pt]
		
		\begin{array}{l}
			\begin{pmatrix}
				|\alpha|^2(1-q_w)^2 & \alpha\beta^*(1-p)(1-q_w)  \\
				\alpha^*\beta(1-p)(1-q_w) & |\beta|^2(1-p)^2
			\end{pmatrix} 
			\otimes
			\begin{pmatrix}
				|\gamma|^2(1-p)^2 & \gamma\delta^*(1-p)(1-q_w)  \\
				\gamma^*\delta(1-p)(1-q_w)  & |\delta|^2(1-q_w)^2
			\end{pmatrix}\\[8pt]
			\hfill j=1,2 \text{ and } i=3,4
		\end{array} \\[12pt]
		
		\begin{array}{l}
			\begin{pmatrix}
				|\alpha|^2(1-p)^2 & \alpha\beta^*(1-p)(1-q_w) \\
				\alpha^*\beta(1-p)(1-q_w)  & |\beta|^2(1-q_w)^2
			\end{pmatrix} 
			\otimes
			\begin{pmatrix}
				|\gamma|^2(1-q_w)^2 & \gamma\delta^*(1-p)(1-q_w) \\
				\gamma^*\delta(1-p)(1-q_w) & |\delta|^2(1-p)^2
			\end{pmatrix}\\[8pt]
			 \hfill j=3,4 \text{ and } i=1,2
		\end{array} \\[12pt]
		
		\begin{array}{l}
			\begin{pmatrix}
				|\alpha|^2(1-p)^2 & \alpha\beta^*(1-p)(1-q_w)  \\
				\alpha^*\beta(1-p)(1-q_w)  & |\beta|^2(1-q_w)^2
			\end{pmatrix}
			\otimes
			\begin{pmatrix}
				|\gamma|^2(1-p)^2 & \gamma\delta^*(1-p)(1-q_w)  \\
				\gamma^*\delta(1-p)(1-q_w)  & |\delta|^2(1-q_w)^2
			\end{pmatrix}\\[8pt]
			 \hfill j,i=3,4
		\end{array}
	\end{cases}
	\label{eout}
\end{eqnarray}

where $g^{'}_{M_j^BM_i^A}=\text{Tr}(\rho^{'}_{M^{B}_j M^{A}_i})$ is the corresponding success probability of BQT, and for different sixteen possible output states given in Eq.~\ref{eout} is
\begin{eqnarray}
	g^{'}_{M^{B}_j M^{A}_i} &=&  
	\begin{cases}
	\frac{1}{4\big(1+(1-p)^2\big)^2} \bigg[\bigg( |\alpha|^2(1-q_w)^2 + |\beta|^2(1-p)^2\bigg)\times \bigg( |\gamma|^2(1-q_w)^2 + |\delta|^2(1-p)^2\bigg) \bigg] & j,i=1,2 \\\\
		
			\frac{1}{4\big(1+(1-p)^2\big)^2} \bigg[\bigg( |\alpha|^2(1-q_w)^2 + |\beta|^2(1-p)^2\bigg)\times \bigg( |\gamma|^2(1-p)^2 + |\delta|^2(1-q_w)^2\bigg) \bigg] &  j=1,2 \text{ and } i=3,4 \\\\
		
			\frac{1}{4\big(1+(1-p)^2\big)^2} \bigg[\bigg( |\alpha|^2(1-p)^2 + |\beta|^2(1-q_w)^2\bigg)\times \bigg( |\gamma|^2(1-q_w)^2 + |\delta|^2(1-p)^2\bigg) \bigg] &  j=3,4 \text{ and } i=1,2 \\\\
		
			\frac{1}{4\big(1+(1-p)^2\big)^2} \bigg[\bigg( |\alpha|^2(1-p)^2 + |\beta|^2(1-q_w)^2\bigg)\times \bigg( |\gamma|^2(1-p)^2 + |\delta|^2(1-q_w)^2\bigg) \bigg]&  j,i=3,4
	\end{cases}
	\label{egbt}
\end{eqnarray}
By setting the weak measurement strength equal to the ADC strength, i.e., $q_w = p$, it follows from Eqs.~\ref{eout} and \ref{egbt} that all sixteen possible teleported states are obtained with equal teleportation success probability and ultimately reduce to the same final state.
 Hence, the total success probability of BQT with all qubits undergoing decoherence is
\begin{eqnarray}
	g_t^{BT-EW'}=\sum_{j,i=1}^{4}	g^{'}_{M^{B}_j M^{A}_i}=\bigg(1-\frac{(2q_w-q_w^2)}{1+(1-p)^2}\bigg)^2
	\label{egt}
\end{eqnarray}
Furthermore, the fidelity between the output states given in Eq.~\ref{eout} and the input states in Eq.~\ref{inp} for both partners follows the same definition as in Eq.~\ref{fid} but in the present scenario takes the form  
\begin{eqnarray}
	F^{B'A'}_{ji} &=& \text{Tr} \bigg(\rho_a \otimes \rho_b \, \rho^{'}_{M^{B}_j M^{A}_i} \bigg) = 	F_j^{B'} F_i^{A'} , \quad \text{where}\nonumber \\
	F_j^{B'} F_i^{A'} &=&
	\begin{cases}
		\Bigg( \frac{(|\beta|^2p + |\alpha|^2q_w-1)^2}{|\alpha|^2(q_w^2-2q_w) + |\beta|^2(p^2-2p)+1} \Bigg)
		\Bigg( \frac{(|\delta|^2p + |\gamma|^2q_w-1)^2 }{|\gamma|^2(q_w^2-2q_w) + |\delta|^2(p^2-2p)+1} \Bigg) & j,i=1,2 \\\\[10pt]
		
		\Bigg( \frac{(|\beta|^2p + |\alpha|^2q_w-1)^2}{|\alpha|^2(q_w^2-2q_w) + |\beta|^2(p^2-2p)+1} \Bigg)
	\Bigg( \frac{(|\delta|^2q_w + |\gamma|^2p-1)^2 }{|\gamma|^2(p^2-2p) + |\delta|^2(q_w^2-2q_w)+1} \Bigg)&  j=1,2 \text{ and } i=3,4 \\\\[10pt]
		
		\Bigg( \frac{(|\beta|^2q_w+ |\alpha|^2p-1)^2}{|\alpha|^2(p^2-2p) + |\beta|^2(q_w^2-2q_w)+1} \Bigg)
	\Bigg( \frac{(|\delta|^2p + |\gamma|^2q_w-1)^2 }{|\gamma|^2(q_w^2-2q_w) + |\delta|^2(p^2-2p)+1} \Bigg)  & j=3,4 \text{ and } i=1,2 \\\\[10pt]
		
		\Bigg( \frac{(|\beta|^2q_w+ |\alpha|^2p-1)^2}{|\alpha|^2(p^2-2p) + |\beta|^2(q_w^2-2q_w)+1} \Bigg)
	\Bigg( \frac{(|\delta|^2q_w + |\gamma|^2p-1)^2 }{|\gamma|^2(p^2-2p) + |\delta|^2(q_w^2-2q_w)+1} \Bigg) &  j,i=3,4
	\end{cases}
	\label{efid}
\end{eqnarray}
and upon considering the probabilities of sixteen possible projections given in Eq.~\ref{epro}, the total fidelity is defined in a similar manner as Eq.~\ref{tf}, and is given by
\begin{eqnarray}
	\sum_{j,i=1}^{4}P_i^{A'}P_j^{B'}F_{ji}^{B'A'}
	\label{etf}
\end{eqnarray}
Now, to quantify the performance of the BQT protocol when entire four qubit channel is subjected to ADC effects, and to observe the effectiveness of weak measurements defined in Eq.~\ref{ewm} in suppressing the noise, average teleportation fidelity is calculated as
\begin{eqnarray}
	F_{av}^{BT-EW'}=\int_{0}^{1}\sum_{j,i=1}^{4}P_i^{A'}P_j^{B'}F_{ji}^{B'A'} d|\alpha|^2
	\label{efav}
\end{eqnarray} 
Once again, we assume that Alice's and Bob's input states are arbitrary but the same, i.e., $\alpha = \gamma$ and $\beta = \delta$, as discussed in Sec.~\ref{sub1}. It is observed that when the entire four-qubit entangled state given by Eq.~ \ref{eqtch1}, is passed through independent ADCs with the same decay rate $p$, the entanglement between Alice's sub-system marked by qubits $(1,3)$ and Bob's sub-system $(2,4)$ is reduced compared to the scenario where only the recovery qubits $(2,3)$ are subjected to ADC. This is described by observing the behaviour of Von-Neumann entropy shown in Fig.~\ref{von} and discussed in Appendix \ref{appA}.  This decrease in entanglement results in BQT with reduced average fidelity than that of Sec.~\ref{sub1}, as illustrated in the graphical discussion below.
%\clearpage
\subsubsection{Numerical Results and Analysis}
\begin{figure}[H]
	\centering
	\begin{subfigure}[b]{0.45\textwidth}
		\includegraphics[width=9cm, height=6.5cm]{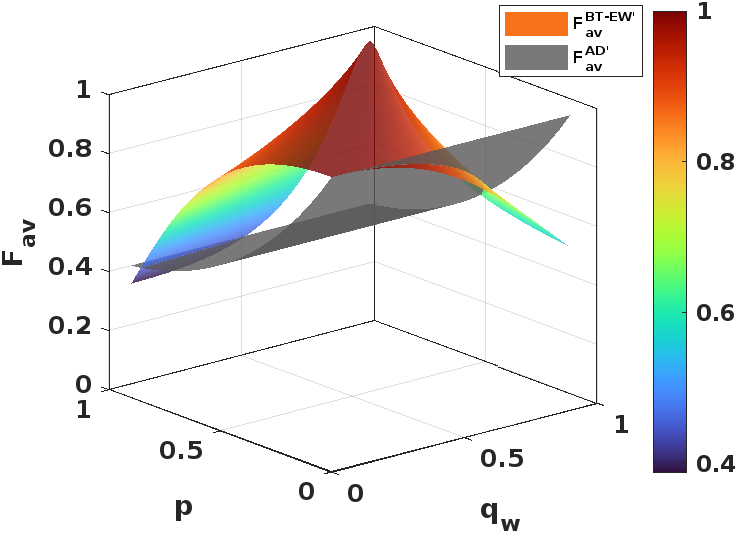}
		\caption{}
		\label{eF3d}
	\end{subfigure}
	\hspace{0.05\textwidth}
	\begin{subfigure}[b]{0.45\textwidth}
		\includegraphics[width=\textwidth, height=6cm]{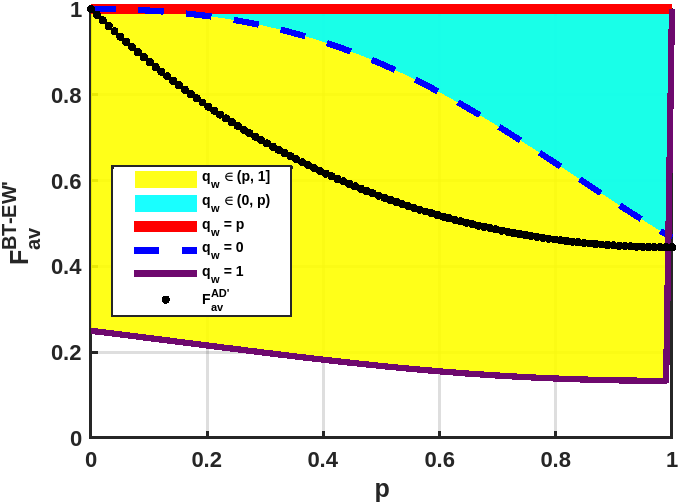}
		\caption{}
		\label{eF2d}
	\end{subfigure}
	\caption{\ref{eF3d}  Average fidelity $F_{av}^{BQT-EW'}$, when all qubits of the entangled state are subjected to ADCs, as a function of the decoherence rate $p$ and weak measurement strength $q_w$. The red surface illustrates the enhanced fidelity with protection, while the gray plane represents the average fidelity $F_{av}^{AD'}$ of BQT in the presence of AD without protection. \ref{eF2d} Comparison of average fidelity $F_{av}^{BT-EW'}$ as a function of decoherence rate $p$ and weak measurement strength $q_w$. The cyan region represents $q_w \in (0, p)$, while the yellow region indicates $q_w \in (p, 1]$. The red line shows the fidelity for $q_w = p$, the blue dotted line for $q_w = 0$, and the purple line for $q_w = 1$. The black dots represent the average fidelity $F_{av}^{AD'}$ without protection.}
	\label{efiall}
\end{figure}

\begin{figure}[H]
	\centering
	\begin{subfigure}[b]{0.45\textwidth}
		\includegraphics[width=9cm, height=6cm]{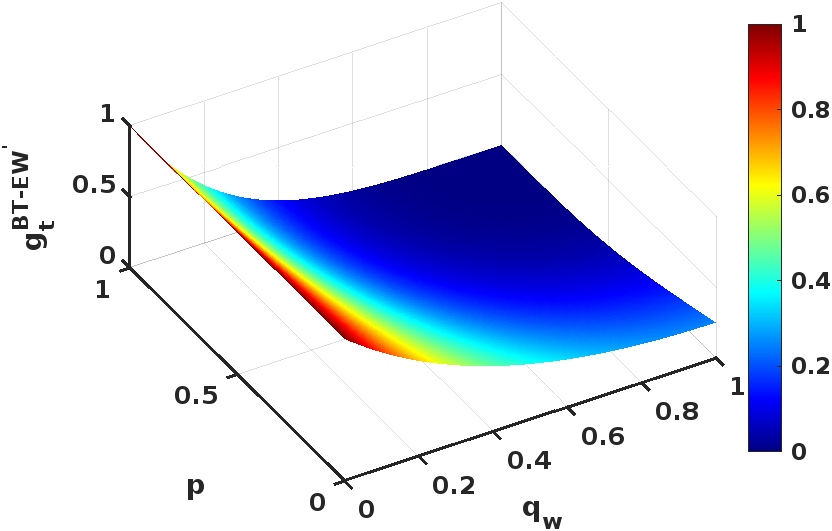}
		\caption{}
		\label{egt3d}
	\end{subfigure}
	\hspace{0.05\textwidth}
	\begin{subfigure}[b]{0.45\textwidth}
		\includegraphics[width=8cm, height=6cm]{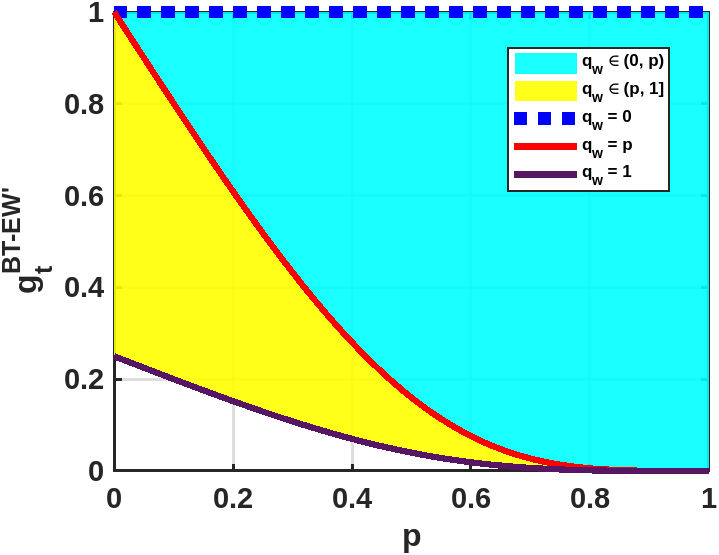}
		\caption{}
		\label{egt2d}
	\end{subfigure}
\caption{(\ref{egt3d}) Total teleportation success probability of the proposed BQT protocol, considering all qubits of the entangled channel are subject to ADC, as a function of the weak measurement strength $q_w$ and decay rate $p$. (\ref{egt2d}) Comparison of total teleportation success probability with respect to $p$ for different value of $q_w$. The cyan shaded region represents $g_t^{BT-EW'}$ for $q_w \in (0, p)$, the yellow shaded region $q_w\in(p,1]$ represents strictly prohibited $q_w$ values, the red solid line corresponds to $q_w = p$, purple solid line $q_w=1$ and the blue dashed line corresponds to $q_w = 0$.}
	\label{egall}
\end{figure}

Here, the effectiveness of our optimization strategy is illustrated by analyzing the behavior of the average fidelity, given in Eq.~\ref{efav} and graphically represented in Fig.~\ref{efiall}. Additionally, the total success probability of BQT, described in Eq.~\ref{egt}, is depicted in Fig.~\ref{egall}. For comparison, we have also considered the average fidelity of the unprotected BQT scheme discussed in Appendix \ref{apC}. This is represented by the gray plane in Fig.~\ref{eF3d} and the black dotted line in Fig.~\ref{eF2d}, and is given by
\begin{eqnarray}
	F_{av}^{AD'}	=	\frac{1}{9}\bigg(3+p^2-2p\bigg)^2
	\label{eunp}
\end{eqnarray} 

When the entire four qubit channel is subjected to ADC, the noisy effects reduce the entanglement between the subsystems of Alice and Bob as compared to case when only recovery qubits are under ADC as observed clearly from  the behaviour of Von Neumann Entropy in Fig.~\ref{von}. This reduction in entanglement has a direct consequence on the behaviour of average fidelity shown in Fig.~\ref{efiall}, it is observed by comparing Figs.\ref{F2d} and \ref{eF2d}, that  $F_{av}^{BT-EW}$ is greater than $F_{av}^{BT-EW'}$ at all ADC and weak measurement strengths. It is more subtle by observing the behaviour of average fidelity in absence of any weak measurements i.e, $q_w=0$ shown by the blue dashed line. It is clear that EAM are more effective in channel protection for the case discussed in Sec.~\ref{sub1} than Sec.~\ref{sub2}. In Fig.~\ref{eF3d}, again we show that our protected BQT scheme occurs with higher fidelity (red plane) than corresponding unprotected BQT (gray plane) in presence of ADC. Also in Fig.~\ref{eF2d}, the results show that BQT is again optimized for $q_w=p$ marked by red solid line with  $F_{av}^{BT-EW'}=1$. By increasing the weak measurement strength $q_w$, a higher $F_{av}^{BT-EW'}$ is obtained at the cost of reduced $g_t^{BT-EW'}$. However, by tuning $q_w\in(0,p)$ a perfect balance between $F_{av}^{BT-EW'}$ and  $g_t^{BT-EW'}$ is achieved shown by the cyan shaded region in Figs.~\ref{eF2d} and \ref{egt2d}. Moreover, it is observed that $q_w\in(p,1]$, depicted by the yellow shaded region is prohibited because of lower $F_{av}^{BT-EW'}$ and  $g_t^{BT-EW'}$ than at $q_w=p$.
%\vspace{6cm}
%\clearpage
\section{conclusion}
\label{sec3}
In conclusion, this paper proposes strategies  to optimize the fidelity of quantum information transfer between two nodes in the presence of decoherence, by protecting the quantum channel utilizing EAM and weak measurements. To add additional layer of security to our proposed BQT scheme a third party supervises the protocol during the entanglement distribution process. The  protocol involves (a) Preparation of four-qubit entangled channel. (b)  Entanglement distribution by third party through ADCs.  (c) Protected BQT with weak measurements.  The protocol is explicitly discussed for ADC noisy effects, however it works well for all decoherence channels having a minimum of one invertible Kraus operator. We have analyzed two scenarios~ \ref{sub1} BQT with Amplitude Damping on Recovery Qubits ($2$ and$3$) and~ \ref{sub2} BQT with Amplitude Damping on all Qubits. The quantum circuits for both scenarios are shown in Figs.~\ref{rbqt} and \ref{ebqt} demonstrating the entire optimized BQT protocol.
 By analyzing these two  scenarios, it is observed that the proposed scheme achieves a balance between average fidelity and success probability when the weak measurement strength is bounded by the decay rate i.e, $q_w\in(0,p)$. Moreover, when the weak measurement strength is set equal to the ADC strength, i.e., $q_w = p$, the protected BQT scheme achieves unit fidelity and a non-zero total success probability. This condition signifies complete suppression of decoherence, independent of the ADC strength.
  Since the weak measurement strength governs the optimization of the proposed protocol, we observe that fidelity and success probability decrease when it exceeds the ADC strength. This defines the prohibited domain $q_w \in (p,1]$
   and  highlights the importance of carefully tuning this parameter to counteract decoherence. Among the two cases, the case \ref{sub1} where only recovery qubits are passed through ADCs performs better by maintaining higher entanglement than case \ref{sub2} when entire four qubit channel is subjected to ADC effects. This is demonstrated by observing the behaviour of Von-Neumann entropy shown in Fig.~\ref{von} and also by the behaviour of average fidelity for the respective cases. For comparative analysis the unprotected BQT schemes are discussed in Appendices \ref{apB} and \ref{apC}. It is observed from the  behaviour of average fidelity in Figs.~\ref{fiall} and \ref{efiall}, that our optimized protocol BT-EW outperforms the unprotected scheme in both cases.  Overall, this study introduces the first protected BQT scheme for bilateral quantum information transfer, paving the way for the realization of reliable distributed quantum networks. The findings of this paper can be extended to other types of decoherence channels having one invertible Kraus operator. Moreover, the proposed scheme identifies the weak measurement strength values effective in suppression of the decoherence.
   \vspace{2cm}
    
   \noindent\textbf{Funding:} We have not received any funds for this work. \\
   
   \noindent\textbf{Data availability:}  No data are associated with this manuscript. \\

 \newpage
 \appendix
\section{BQT with Amplitude damping on recovery qubits (2 and 3) without protection}
\label{apB}
For comparative analysis we study the BQT protocol when  only recovery qubits are subjected to ADC effects without EAM and weak measurements. Average fidelity is computed and the absence of EAM and weak measurements makes it deterministic with unit success probability. The four qubit entangled state without discarding measurements outcomes during entanglement distribution is given by.
\begin{eqnarray}
	\rho_{1234}^{f^{AD}} &=& \frac{K_i^A K_j^B \rho_{1234} K_j^{B\dagger} K_i^{A\dagger}}{\text{Tr}(K_i^A K_j^B \rho_{1234} K_j^{B\dagger} K_i^{A\dagger})} 
\end{eqnarray}
\begin{equation}
	\scriptsize
	\rho_{\text{1234}}^{f^{AD}} = \frac{1}{4} 
	\left(
	\begin{array}{cccccccccccccccc}
		1 & 0 & 0 & \sqrt{1-p} & 0 & 0 & 0 & 0 & 0 & 0 & 0 & 0 & \sqrt{1-p} & 0 & 0 & (1-p) \\
		0 & 0 & 0 & 0 & 0 & 0 & 0 & 0 & 0 & 0 & 0 & 0 & 0 & 0 & 0 & 0 \\
		0 & 0 & p & 0 & 0 & 0 & 0 & 0 & 0 & 0 & 0 & 0 & 0 & 0 & p\sqrt{1-p} & 0 \\
		\sqrt{1-p} & 0 & 0 & (1-p) & 0 & 0 & 0 & 0 & 0 & 0 & 0 & 0 & (1-p) & 0 & 0 & (1-p)\sqrt{1-p} \\
			0 & 0 & 0 & 0 & 0 & 0 & 0 & 0 & 0 & 0 & 0 & 0 & 0 & 0 & 0 & 0 \\
				0 & 0 & 0 & 0 & 0 & 0 & 0 & 0 & 0 & 0 & 0 & 0 & 0 & 0 & 0 & 0 \\
					0 & 0 & 0 & 0 & 0 & 0 & 0 & 0 & 0 & 0 & 0 & 0 & 0 & 0 & 0 & 0 \\
						0 & 0 & 0 & 0 & 0 & 0 & 0 & 0 & 0 & 0 & 0 & 0 & 0 & 0 & 0 & 0 \\
		0 & 0 & 0 & 0 & 0 & 0 & 0 & 0 & p & 0 & 0 & p\sqrt{1-p} & 0 & 0 & 0 & 0 \\
			0 & 0 & 0 & 0 & 0 & 0 & 0 & 0 & 0 & 0 & 0 & 0 & 0 & 0 & 0 & 0 \\
				0 & 0 & 0 & 0 & 0 & 0 & 0 & 0 & 0 & 0 & p^2 & 0 & 0 & 0 & 0 & 0 \\	
					0 & 0 & 0 & 0 & 0 & 0 & 0 & 0 & p\sqrt{1-p} & 0 & 0 & p(1-p) & 0 & 0 & 0 & 0 \\	
	\sqrt{1-p} & 0 & 0 & (1-p) & 0 & 0 & 0 & 0 & 0 & 0 & 0 & 0 & (1-p) & 0 & 0 & (1-p)\sqrt{1-p} \\
		0 & 0 & 0 & 0 & 0 & 0 & 0 & 0 & 0 & 0 & 0 & 0 & 0 & 0 & 0 & 0 \\
			0 & 0 & p\sqrt{1-p} & 0 & 0 & 0 & 0 & 0 & 0 & 0 & 0 & 0 & 0 & 0 & p(1-p) & 0 \\	
			(1-p) & 0 & 0 & (1-p)\sqrt{1-p} & 0 & 0 & 0 & 0 & 0 & 0 & 0 & 0 & (1-p)\sqrt{1-p} & 0 & 0 & (1-p)^2 \\							
	\end{array}
	\right)
	\label{chap1}
\end{equation}
with $\text{Tr}	(\rho_{\text{1234}}^{f^{AD}}) =1$, which is attributed to the absence of EAM and retention of both $k_0$ and $k_1$ measurement outcomes. After following similar steps as in protected protocol except application of weak measurements in the last step, the teleported states post unitary operations  $U_j^BU_i^A $  with $U_i^A, ~U_j^B= I (i,j=1), \sigma_z (i,j=2), \sigma_x (i,j=3)$ and $\sigma_x \sigma_z (i,j=4)$ takes the form
\begin{eqnarray}
	\rho^{AD}_{U_j U_i} &=\frac{1}{16g^{AD}_{U_jU_i}}
	\begin{cases}
		\begin{array}{l}
			\begin{pmatrix}
				|\alpha|^2+p|\beta|^2 & \alpha\beta^*\sqrt{1-p} \\
				\alpha^*\beta\sqrt{1-p} & |\beta|^2(1-p)
			\end{pmatrix}
			\otimes
			\begin{pmatrix}
				|\gamma|^2+p|\delta|^2 & \gamma\delta^*\sqrt{1-p} \\
				\gamma^*\delta\sqrt{1-p} & |\delta|^2(1-p)
			\end{pmatrix} ~~
			\hfill j,i=1,2
		\end{array} \\[12pt]
		
		\begin{array}{l}
			\begin{pmatrix}
			|\alpha|^2+p|\beta|^2 & \alpha\beta^*\sqrt{1-p} \\
			\alpha^*\beta\sqrt{1-p} & |\beta|^2(1-p)
		\end{pmatrix}
		\otimes
		\begin{pmatrix}
			|\gamma|^2(1-p) & \gamma\delta^*\sqrt{1-p} \\
			\gamma^*\delta\sqrt{1-p} &|\gamma|^2p+ |\delta|^2
		\end{pmatrix} ~~
			\hfill j=1,2 \text{ and } i=3,4
		\end{array} \\[12pt]
		
		\begin{array}{l}
			\begin{pmatrix}
			|\alpha|^2(1-p) & \alpha\beta^*\sqrt{1-p} \\
			\alpha^*\beta\sqrt{1-p} &|\alpha|^2p+ |\beta|^2
		\end{pmatrix}
		\otimes
		\begin{pmatrix}
			|\gamma|^2+p|\delta|^2 & \gamma\delta^*\sqrt{1-p} \\
			\gamma^*\delta\sqrt{1-p} & |\delta|^2(1-p)
		\end{pmatrix} ~~
			\hfill j=3,4 \text{ and } i=1,2
		\end{array} \\[12pt]
		
		\begin{array}{l}
			\begin{pmatrix}
			|\alpha|^2(1-p) & \alpha\beta^*\sqrt{1-p} \\
			\alpha^*\beta\sqrt{1-p} &|\alpha|^2p+ |\beta|^2
		\end{pmatrix}
		\otimes
		\begin{pmatrix}
			|\gamma|^2(1-p) & \gamma\delta^*\sqrt{1-p} \\
			\gamma^*\delta\sqrt{1-p} &|\gamma|^2p+ |\delta|^2
		\end{pmatrix} ~~
			\hfill j,i=3,4
		\end{array}
	\end{cases}
	\label{ap1}
\end{eqnarray}
where $g^{AD}_{U_jU_i}=\frac{1}{16}$ are the probabilities of obtaining each output state $\rho^{AD}_{U_j U_i}$ by the two partners and is also equal to the joint probability $P^{AD}_{ji}$ of measuring the bell states. Now the fidelity corresponding to output states in Eq.~\ref{ap1} is

\begin{eqnarray}
	f^{AD}_{ji} &=&  
	\begin{cases}
		\begin{array}{l}
		\bigg( |\alpha|^4+ |\beta|^4(1-p)+|\alpha|^2|\beta|^2(p+2\sqrt{1-p})\bigg)\bigg( |\gamma|^4+ |\delta|^4(1-p)+|\gamma|^2|\delta|^2(p+2\sqrt{1-p})\bigg)\\
			\hfill j,i=1,2
		\end{array} \\[12pt]
		
		\begin{array}{l}
			\bigg( |\alpha|^4+ |\beta|^4(1-p)+|\alpha|^2|\beta|^2(p+2\sqrt{1-p})\bigg)\bigg( |\delta|^4+ |\gamma|^4(1-p)+|\gamma|^2|\delta|^2(p+2\sqrt{1-p})\bigg) \\[8pt] 
			\hfill j=1,2 \text{ and } i=3,4
		\end{array} \\[12pt]
		
		\begin{array}{l}
		\bigg( |\beta|^4+ |\alpha|^4(1-p)+|\alpha|^2|\beta|^2(p+2\sqrt{1-p})\bigg)\bigg( |\gamma|^4+ |\delta|^4(1-p)+|\gamma|^2|\delta|^2(p+2\sqrt{1-p})\bigg) \\[8pt] 
			\hfill j=3,4 \text{ and } i=1,2
		\end{array} \\[12pt]
		
		\begin{array}{l}
			\bigg( |\beta|^4+ |\alpha|^4(1-p)+|\alpha|^2|\beta|^2(p+2\sqrt{1-p})\bigg)\bigg( |\delta|^4+ |\gamma|^4(1-p)+|\gamma|^2|\delta|^2(p+2\sqrt{1-p})\bigg)  \\[8pt] 
			\hfill j,i=3,4
		\end{array}
	\end{cases}
	\label{fidap1}
\end{eqnarray} 
Hence, the average fidelity considering the fact that $g^{AD}_{U_jU_i}=P^{AD}_{ji}=\frac{1}{16};(j,i=1,2,3,4)$, also setting $\alpha=\gamma$ and $\beta=\delta$ is

\begin{eqnarray}
	F_{av}^{AD}=	\frac{1}{9}\bigg(2-\frac{p}{2}+\sqrt{1-p}\bigg)^2
	\label{avap1}
\end{eqnarray} 
\section{BQT with Amplitude damping on all qubits without protection}
\label{apC}
Here, we compute the average fidelity when the entire four qubit channel is subjected to decoherence without protection shown by the grey plane in Fig.~\ref{eF3d}. The four qubit channel without protection takes the form
\begin{eqnarray}
	\rho_{1234}^{f^{AD'}} &=& \frac{K_{ii'}^A K_{jj'}^B \rho_{1234} K_{jj'}^{B\dagger} K_{ii'}^{A\dagger}}{\text{Tr}(K_{ii'}^A K_{jj'}^B \rho_{1234} K_{jj'}^{B\dagger} K_{ii'}^{A\dagger})}
\end{eqnarray}
\begin{equation}
		\rho_{\text{1234}}^{f^{AD'}} = \frac{1}{4} 
		\begin{pmatrix}
		(1+p^2)&0&0&(1-p)\\
		0&p(1-p)&0&0\\
		0&0&p(1-p)&0\\
		(1-p)&0&0&(1-p)^2
		\end{pmatrix}\otimes\begin{pmatrix}
		(1+p^2)&0&0&(1-p)\\
		0&p(1-p)&0&0\\
		0&0&p(1-p)&0\\
		(1-p)&0&0&(1-p)^2
		\end{pmatrix}
	\label{chap2}
\end{equation}
again $\text{Tr}	(\rho_{\text{1234}}^{f^{AD'}}) =1$, due to absence of EAM during entanglement distribution process. The teleported states after the recovery operations by the two partners takes the form 

\begin{eqnarray}
	\scriptsize
	\rho^{AD'}_{U_j U_i} &=\frac{1}{16g^{AD'}_{U_jU_i}}
	\begin{cases}
		\begin{array}{l}
			\footnotesize
			\begin{pmatrix}
				|\alpha|^2(1+p^2)+|\beta|^2p(1-p) & \alpha\beta^*(1-p) \\
				\alpha^*\beta(1-p) &|\alpha|^2p(1-p) +|\beta|^2(1-p)^2
			\end{pmatrix}
			\otimes
				\begin{pmatrix}
				|\gamma|^2(1+p^2)+|\delta|^2p(1-p) & \gamma\delta^*(1-p) \\
				\gamma^*\delta(1-p) &|\gamma|^2p(1-p) +|\delta|^2(1-p)^2
			\end{pmatrix} ~~\\
			\hfill j,i=1,2
		\end{array} \\\\[12pt]
		
		\begin{array}{l}
		\footnotesize
		\begin{pmatrix}
			|\alpha|^2(1+p^2)+|\beta|^2p(1-p) & \alpha\beta^*(1-p) \\
			\alpha^*\beta(1-p) &|\alpha|^2p(1-p) +|\beta|^2(1-p)^2
		\end{pmatrix}
		\otimes
		\begin{pmatrix}
			|\gamma|^2(1-p)^2+|\delta|^2p(1-p) & \gamma\delta^*(1-p) \\
			\gamma^*\delta(1-p) &|\gamma|^2p(1-p) +|\delta|^2(1+p^2)
		\end{pmatrix} ~~\\
			\hfill j=1,2 \text{ and } i=3,4
		\end{array} \\\\[12pt]
		
		\begin{array}{l}
		\footnotesize
		\begin{pmatrix}
			|\alpha|^2(1-p)^2+|\beta|^2p(1-p) & \alpha\beta^*(1-p) \\
			\alpha^*\beta(1-p) &|\alpha|^2p(1-p) +|\beta|^2(1+p^2)
		\end{pmatrix}
		\otimes
		\begin{pmatrix}
			|\gamma|^2(1+p^2)+|\delta|^2p(1-p) & \gamma\delta^*(1-p) \\
			\gamma^*\delta(1-p) &|\gamma|^2p(1-p) +|\delta|^2(1-p)^2
		\end{pmatrix} \\~~
			\hfill j=3,4 \text{ and } i=1,2
		\end{array} \\\\[12pt]
		
	\begin{array}{l}
		\footnotesize
		\begin{pmatrix}
			|\alpha|^2(1-p)^2+|\beta|^2p(1-p) & \alpha\beta^*(1-p) \\
			\alpha^*\beta(1-p) &|\alpha|^2p(1-p) +|\beta|^2(1+p^2)
		\end{pmatrix}
		\otimes
		\begin{pmatrix}
			|\gamma|^2(1-p)^2+|\delta|^2p(1-p) & \gamma\delta^*(1-p) \\
			\gamma^*\delta(1-p) &|\gamma|^2p(1-p) +|\delta|^2(1+p^2)
		\end{pmatrix} \\~~
			\hfill j,i=3,4
		\end{array}
	\end{cases}
	\label{ap2}
\end{eqnarray}
with $g^{AD'}_{U_jU_i}=P^{AD'}_{ji}$ is the success probability of obtaining teleportation and is given by  $g^{AD'}_{U_jU_i}=\text{Tr}(\rho^{AD'}_{U_j U_i})$ such that total teleportation success probability $ \displaystyle  \sum_{j,i=1}^4 g^{AD'}_{U_jU_i}=1 $ and  the fidelity with which the input states are teleported is
\begin{eqnarray}
	f^{AD'}_{ji} &=\frac{1}{16P^{AD'}_{ji}}&  
	\begin{cases}
		\begin{array}{l}
			\bigg( |\alpha|^4(1+p^2)+ |\beta|^4(1-p)^2+2|\alpha|^2|\beta|^2(1-p^2)\bigg)	\bigg( |\gamma|^4(1+p^2)+ |\delta|^4(1-p)^2+2|\gamma|^2|\delta|^2(1-p^2)\bigg)\\
			\hfill j,i=1,2
		\end{array} \\[12pt]
		
	\begin{array}{l}
		\bigg( |\alpha|^4(1+p^2)+ |\beta|^4(1-p)^2+2|\alpha|^2|\beta|^2(1-p^2)\bigg)	\bigg( |\gamma|^4(1-p)^2+ |\delta|^4(1+p^2)+2|\gamma|^2|\delta|^2(1-p^2)\bigg) \\[8pt] 
			\hfill j=1,2 \text{ and } i=3,4
		\end{array} \\[12pt]
		
	\begin{array}{l}
		\bigg( |\alpha|^4(1-p)^2+ |\beta|^4(1+p^2)+2|\alpha|^2|\beta|^2(1-p^2)\bigg)	\bigg( |\gamma|^4(1+p^2)+ |\delta|^4(1-p)^2+2|\gamma|^2|\delta|^2(1-p^2)\bigg)\\[8pt] 
			\hfill j=3,4 \text{ and } i=1,2
		\end{array} \\[12pt]
		
		\begin{array}{l}
			\bigg( |\alpha|^4(1-p)^2+ |\beta|^4(1+p^2)+2|\alpha|^2|\beta|^2(1-p^2)\bigg)	\bigg( |\gamma|^4(1-p)^2+ |\delta|^4(1+p^2)+2|\gamma|^2|\delta|^2(1-p^2)\bigg) \\[8pt] 
			\hfill j,i=3,4
		\end{array}
	\end{cases}
	\label{fidap2}
\end{eqnarray} 
Now, by again setting $\alpha=\gamma$ and $\beta=\delta$ the average fidelity of unprotected BQT with all the qubits subjected to ADC effects is 
\begin{eqnarray}
	F_{av}^{AD'}	=	\frac{1}{9}\bigg(3+p^2-2p\bigg)^2
	\label{avap2}
\end{eqnarray} 
\section{Von-Neumann Entropy}
\label{appA}
\begin{figure}[H]
	\centering
	\includegraphics[width=9cm,height=8cm]{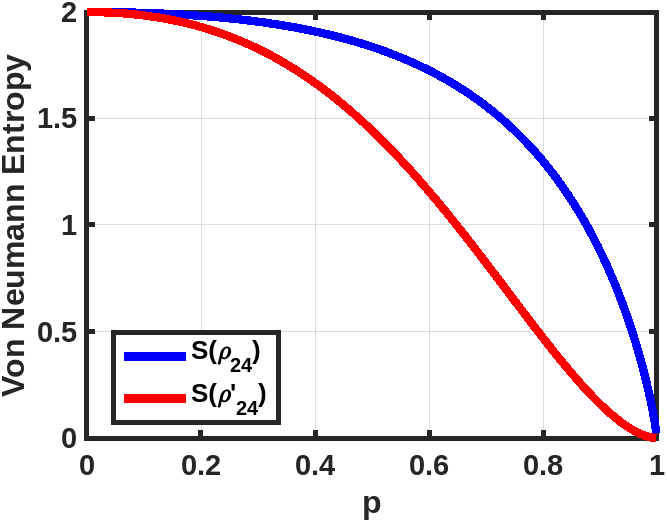}
	\caption{Von Neumann Entropy of reduced density matrix quantifying the entanglement between Alice's (1,3) and Bob's (2,4) subsystems as a function of ADC strength $p$. The blue solid line $S(\rho_{24})$ is Von Neumann entropy when only recovery qubits are passed through ADC and red solid line $S(\rho'_{24})$, when all qubits are subjected to noise. }
	\label{von}
\end{figure}

The Von-Neumann entropy or the entanglement entropy quantifies the degree of entanglement between subsystems of a quantum system and is defined for a quantum state $\rho$ by the formula
\begin{eqnarray}
	S(\rho)=-\text{tr}(\rho \log\rho)
\end{eqnarray}
where logarithms are taken to base two. The entropy is non-negative and in a $d$-dimensional Hilbert space the maximum value is $\log d$. For a composite system divided into subsystems $A$ and $B$ the entanglement entropy $S(\rho_A)$ is calculated as $-\text{Tr}(\rho_A \log_2 \rho_A)$, where $\rho_A$ is the reduced density matrix of $A$. To account for the observed behaviour of average fidelity in above discussed two scenarios \ref{sub1} and \ref{sub2}, Von- Neumann entropy is computed by tracing out Alice's subsystem to  quantify the amount of entanglement between Alice's subsystem marked by qubits $(1,3)$ and Bob's  subsystem $(2,4)$ for both cases and are graphically shown in Fig.\ref{von}.
\begin{eqnarray}
	S(\rho_{24})=-\text{tr}_{13}(\rho^f_{1234} \log\rho^f_{1234})
\end{eqnarray}
\begin{eqnarray}
	S(\rho'_{24})=-\text{tr}_{13}(\rho'^f_{1234} \log\rho'^f_{1234})
\end{eqnarray}
where $S(\rho_{24})$ is the Von-Neumann entropy when only recovery qubits are subjected to ADC marked by blue solid line in Fig.\ref{von} and  $S(\rho'_{24})$ when all four qubits are exposed to ADC by red solid line. Also $ \rho_{1234}$ and $\rho_{\text{1234}}^{f'}$ are defined by Eq.s \ref{chem} and \ref{wchem}. Clearly  $S(\rho_{24})$  is greater than $S(\rho'_{24})$  at all decaying rates $p$ supporting the similar  observed trend for average fidelity.

%\newpage

\end{document}